\RequirePackage{fix-cm}
\documentclass[smallextended]{svjour3}       
\smartqed  
\usepackage{graphicx}

\usepackage{lineno,hyperref}
\usepackage{amssymb,amsmath,latexsym}
\usepackage{mathrsfs}
\usepackage{tipa}
\usepackage{amsfonts}
\usepackage{graphicx}
\usepackage{epstopdf}
\usepackage{subfigure}
\usepackage[title]{appendix}
\usepackage{color}
\usepackage{multirow}
\usepackage{amsbsy}
\usepackage{longtable}
\usepackage{floatrow}
\usepackage{subfigure}
\usepackage{indentfirst}
\usepackage{paralist}
\newtheorem{cor}{Corollary}
\usepackage[font=small]{caption}
\usepackage{algorithm}
\usepackage{algorithmic}
\modulolinenumbers[5]
\usepackage{xcolor}
\usepackage{hyperref} 
\usepackage[paperwidth=8in,paperheight=10in, margin=.875in]{geometry}

\hypersetup{               
	colorlinks=true,                
}

\AtBeginDocument{\hypersetup{pdfborder={0 0 1}}}

\begin{document}
	
	\title{Estimation of block sparsity in compressive sensing
		\thanks{This work was supported by the Swedish Research Council grant (Reg.No. 340-2013-5342) and the Zhejiang Provincial Natural Science Foundation of China under Grant No. LQ21A010003.}
	}
	
	
	\author{Zhiyong Zhou         \and
		Jun Yu 
	}
	
	
	\institute{Zhiyong Zhou \at
		Department of Statistics, Zhejiang University City College, \\ Hangzhou,
		310015, China \\
		\email{zhiyongzhou@zucc.edu.cn}           
		\and
		Jun Yu \at
		Department of Mathematics and Mathematical Statistics, Ume{\aa} University, \\ Ume{\aa},
		901 87, Sweden\\
		\email{jun.yu@umu.se} 
	}
	
	\date{Received: date / Accepted: date}

\maketitle

\begin{abstract}
Explicitly using the block structure of the unknown signal can achieve better reconstruction performance in compressive sensing. Theoretically, an unknown signal with block structure can be accurately recovered from a few number of under-determined linear measurements provided that it is sufficiently block sparse. From the practical point of view, a severe concern is that the block sparse level appears often unknown. In this paper, we introduce a soft measure of block sparsity $k_\alpha(\mathbf{x})=\left(\lVert\mathbf{x}\rVert_{2,\alpha}/\lVert\mathbf{x}\rVert_{2,1}\right)^{\frac{\alpha}{1-\alpha}}$ with $\alpha\in[0,\infty]$, and propose an estimation procedure by using multivariate centered isotropic symmetric $\alpha$-stable random projections. The limiting distribution of the estimator is established. Simulations are conducted to illustrate our theoretical results.
\end{abstract}

\section{Introduction}

Over the past decade, an extensive literature on compressive sensing (CS) has been developed, see the monographs \cite{ek,fr} for a comprehensive view. Formally, one considers the standard CS model, \begin{align}
\mathbf{y}=A\mathbf{x}+\boldsymbol{\varepsilon}, \label{1.1}
\end{align}
where $\mathbf{y}\in\mathbb{R}^{m\times 1}$ is the measurements, $A\in\mathbb{R}^{m\times N}$ is the measurement matrix, $\mathbf{x}\in\mathbb{R}^N$ is the unknown signal, $\boldsymbol{\varepsilon}$ is the measurement error, and $m\ll N$. The goal of CS is to recover the unknown signal $\mathbf{x}$ based on the under-determined measurements $\mathbf{y}$ and the matrix $A$. It is well-known that under the sparsity assumption of the signal and a properly chosen measurement matrix $A$, $\mathbf{x}$ can be reliably recovered from $\mathbf{y}$ by certain algorithms, such as the Basis Pursuit (BP) \cite{cds}, the Orthogonal Matching Pursuit (OMP) \cite{tg}, the Compressive Sampling Matching Pursuit (CoSaMP) \cite{nt} and the Iterative Hard Thresholding (IHT) \cite{bd1}. Specifically, when the sparsity level of the signal $\mathbf{x}$ is $s=\lVert\mathbf{x}\rVert_0=\mathrm{card}\{j:x_j\neq 0\}$, if $m\geq Cs\ln(N/s)$ with some universal constant $C$, and $A$ is a subgaussian random matrix, then accurate recovery can be guaranteed with high probability.

The sparsity level parameter $\lVert \mathbf{x}\rVert_0$ plays a fundamental role in CS, as the required number of measurements, the properties of measurement matrix $A$, and even some recovery algorithms are all depending on it. However, the sparsity level of a signal is usually unknown in practice. To fill the gap between theory and practice, \cite{l1,l2} proposed a numerically stable measure of sparsity $s_\alpha(\mathbf{x})=\left(\lVert\mathbf{x}\rVert_{\alpha}/\lVert\mathbf{x}\rVert_{1}\right)^{\frac{\alpha}{1-\alpha}}$ with $\alpha\in[0,\infty]$,
which is in ratios of norms. By random linear projections using independent and identically distributed (i.i.d.) univariate symmetric $\alpha$-stable random variables, the author constructed an estimation equation for $s_{\alpha}(\mathbf{x})$ with $\alpha\in(0,2]$ by adopting the characteristic function method and obtained the asymptotic normality of the estimator.

As a natural extension of the sparsity with non-zero entries arbitrarily spread throughout the signal, a sparse signal can exhibit additional structure where the non-zero entries occur in clusters. Such signals are referred to as block sparse \cite{de,ekb,em}. Block sparse signals appear in many practical situations, such as when dealing with multi-band signals \cite{me2}, in measurements of gene expression levels \cite{pvmh}, and in color imaging \cite{mw}. Moreover, block sparse model can be used to treat the problems of multiple measurement vector \cite{ch,crek,em,me1} and sampling signals that lie in a union of subspaces \cite{bd2,em,me2}. The corresponding block sparse signal recovery algorithms have been developed to make explicit use of the block structure to achieve better reconstruction performance, such as the mixed $\ell_2/\ell_1$ norm recovery algorithm \cite{ekb,em,sph}, the mixed $\ell_q/\ell_1$ ($q\geq 1$) norm recovery algorithm \cite{ev}, group lasso \cite{yl} or adaptive group lasso \cite{lbw}, iterative reweighted $\ell_2/\ell_1$ recovery algorithms \cite{zb}, the block version of OMP algorithm \cite{ekb} and the model-based CS algorithms \cite{bcdh}. 

\subsection{Motivations}

The block sparsity level $\lVert \mathbf{x}\rVert_{2,0}$ introduced in this paper plays the same central role in block sparse signal recovery as its counterpart $\lVert \mathbf{x}\rVert_0$ does in non-block sparse signal recovery. In what follows, as discussed in \cite{l2} we illustrate the importance of accurately estimating block sparsity from three aspects: the required number of measurements, the measurement matrix and the recovery algorithms. In addition, we demonstrate an example for a better understanding.
\begin{enumerate}[(a)]
	\item  As we know, there are many algorithms that can reliably recover block sparse signal $\mathbf{x}$ when the number of measurements $m$ is larger than $O(\lVert \mathbf{x}\rVert_{2,0}\ln(N))$. To keep the required number of measurements as small as possible, an accurate block sparsity estimator $\widehat{\lVert \mathbf{x}\rVert_{2,0}}$ is desired.
	\item  An upper bound of $\lVert \mathbf{x}\rVert_{2,0}$ (say $k$) is explicitly related to the incoherent properties of the measurement matrix $A$ such as the block restricted isometry property of order $k$ \cite{em}, and the block null space property of order $k$ \cite{zhou2017recovery}, which can lead to a successful recovery for block $k$-sparse signal. To verify whether a given measurement matrix satisfies these properties, we have to know the $k$ in advance, hence estimating $\lVert \mathbf{x}\rVert_{2,0}$ will be of great importance. 
	\item The block sparsity level of the signal of interest $\mathbf{x}$ is often acting as a tuning parameter in recovery algorithms. For instance, the block version of OMP algorithm \cite{ekb} converges in at most $k$ steps if $k$ is an upper bound of $\lVert \mathbf{x}\rVert_{2,0}$, and the block sparsity level of $\mathbf{x}$ should be used to tune the regularization parameter $\lambda$ in the group lasso algorithm \cite{yl}. Generally, a good estimate of $\lVert \mathbf{x}\rVert_{2,0}$ will enable us to reduce computation time by restricting the possible choices of $k$ or $\lambda$, and guarantee that the reconstructed signal's block sparsity level conforms to the true one. 
	
\end{enumerate}

\begin{figure}[htbp]
	\centering
	\includegraphics[width=\textwidth,height=0.4\textheight,keepaspectratio]{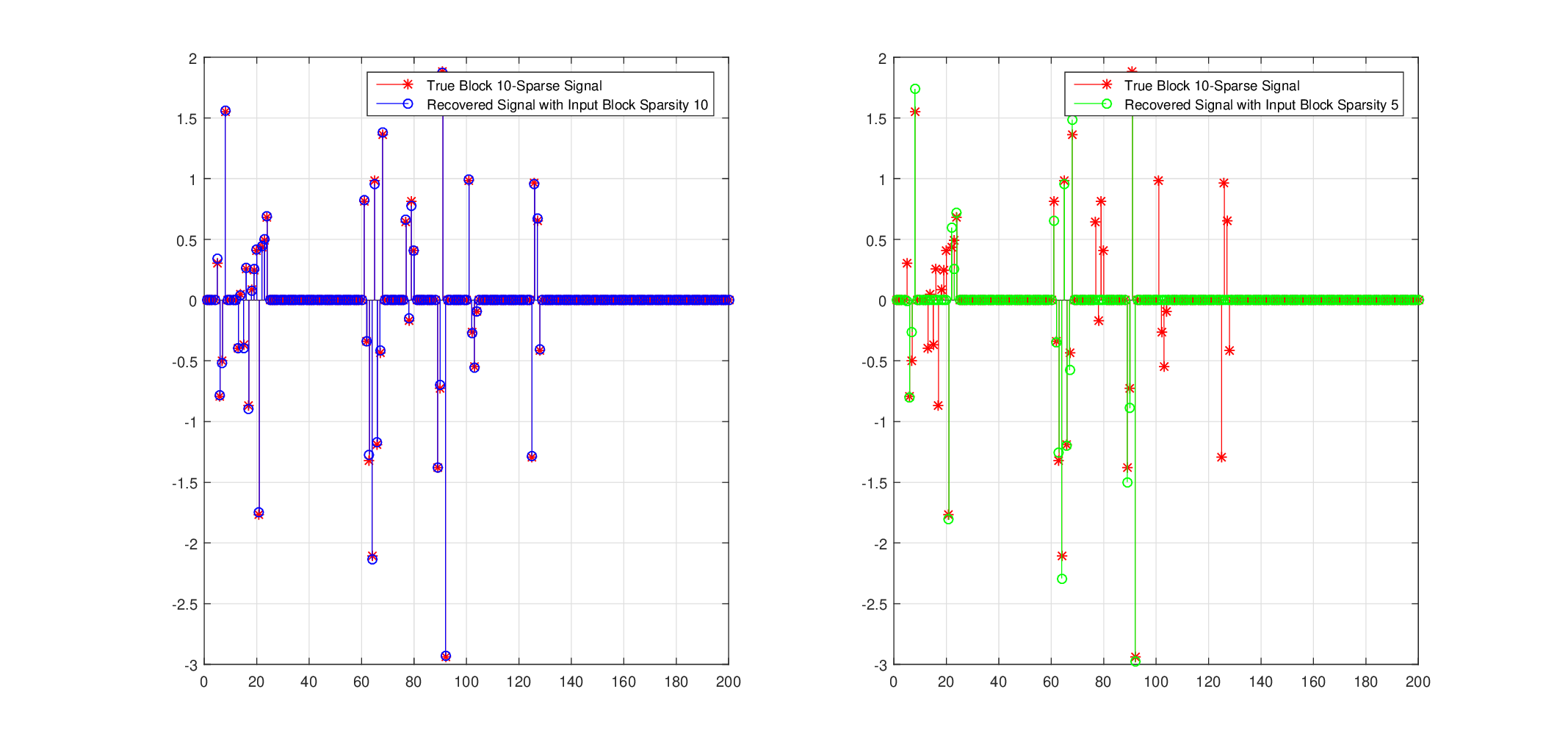}
	\caption{A numerical test for the model-based CoSaMP algorithm with different input block sparsities, left panel (with input block sparsity $10$) and right panel (with input block sparsity $5$)}
	\label{Input_Block_Sparsity}
\end{figure}

\begin{figure}[htbp]
	\centering
	\includegraphics[width=\textwidth,height=0.4\textheight,keepaspectratio]{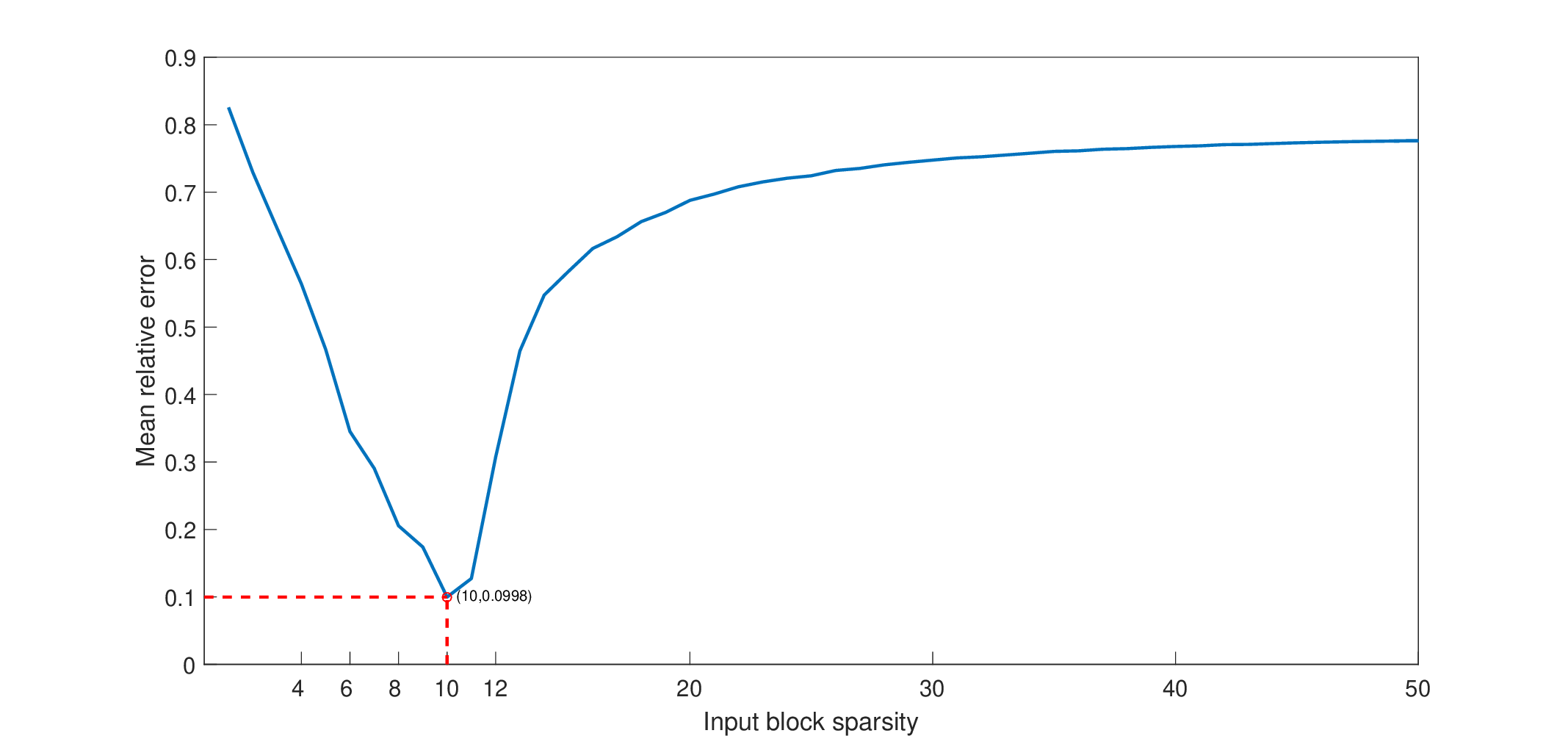}
	\caption{The mean relative error while varying the input block sparsity from $1$ to $50$. The mean relative error takes a minimum value of $0.0998$ at the input block sparsity $10$.}
	\label{Mean_Relative_Error}
\end{figure}

Next, to further demonstrate that the unknown block sparsity level $\lVert \cdot\rVert_{2,0}$ plays a vital role in the block sparse signal recovery algorithms, we present here a test example using the model-based CoSaMP algorithm in \cite{bcdh} for a block $10$-sparse signal $\mathbf{x}\in\mathbb{R}^{200}$ reconstruction with a Gaussian random measurement matrix $A\in\mathbb{R}^{80\times 200}$. The block $10$-sparse signal is generated by choosing $10$ blocks uniformly at random, and then choosing the non-zero entries from the standard normal distribution for these $10$ blocks. We fix the block size $d=4$ and consider the case that the measurements are noise free. As shown in Figure \ref{Input_Block_Sparsity}, a perfect recovery can be achieved via the model-based CoSaMP algorithm with an accurate input block sparsity $10$, while it causes a huge recovery bias if $5$ is used as input. Furthermore, we replicated the experiments $100$ times with different Gaussian random measurement matrices $A$ and evaluated the recovery performance in terms of the mean relative error (MRE) $\frac{\lVert \hat{\mathbf{x}}-\mathbf{x}\rVert_2}{\lVert \mathbf{x}\rVert_2}$. Figure \ref{Mean_Relative_Error} shows how the MRE varies with input block sparsity. We can clearly observe that MRE reaches its minimum when the input block sparsity is exactly the underlying true block sparsity $10$, and it increases as the input block sparsity slips away from the true value.

As a result, obtaining an accurate estimate for the block sparsity level in CS is critical from both theoretical and practical standpoints, and hence it is desirable to develop a new block sparsity estimation approach.

\subsection{Contributions}

In this paper, we first introduce a new soft measure of block sparsity, that is $k_\alpha(\mathbf{x})=\left(\lVert\mathbf{x}\rVert_{2,\alpha}/\lVert\mathbf{x}\rVert_{2,1}\right)^{\frac{\alpha}{1-\alpha}}$ with $\alpha\in[0,\infty]$, and study its properties. It is an extension of the entropy-based non-block sparsity measure $s_\alpha(\mathbf{x})=\left(\lVert\mathbf{x}\rVert_{\alpha}/\lVert\mathbf{x}\rVert_{1}\right)^{\frac{\alpha}{1-\alpha}}$ given in \cite{l1,l2}.

Then we propose an estimator for the block sparsity by using multivariate centered isotropic symmetric $\alpha$-stable random projections and present its asymptotic properties. It covers the results from \cite{l2} as a special case when the block size equals to one (i.e. there is no block structure in the signal).

Finally, a series of numerical experiments are conducted to implement the proposed method and illustrate our theoretical results. To the best of our knowledge, we are the first to address the issue of estimating the block sparsity in CS.

\subsection{Organization and Notations}

The remainder of the paper is organized as follows. In Section 2, we introduce the definition of block sparsity and a soft measure of block sparsity. In Section 3, we present the estimation procedure for the proposed block sparsity measure and obtain the asymptotic properties for the estimators. In Section 4, we conduct simulations to illustrate the theoretical results. Section 5 is devoted to the conclusion. Finally, the proofs are postponed to the Appendix.

Throughout the paper, we denote vectors by boldface lower case letters e.g., $\mathbf{x}$, and matrices by upper case letters e.g., $A$. Vectors are columns by default. $\mathbf{x}^T$ is the transpose of the vector $\mathbf{x}$. The notation $x_j$ denotes the $j$-th component of $\mathbf{x}$. For any vector $\mathbf{x}\in\mathbb{R}^N$, we denote the $\ell_p$-norm  $\lVert\mathbf{x}\rVert_p=(\sum_{j=1}^N|x_j|^p)^{1/p}$ for $p>0$. $I(\cdot)$ is the indicator function. $E$ is the expectation function. $\lfloor\cdot\rfloor$ is the bracket function, which takes the maximum integer value. $\mathrm{Re}(\cdot)$ is the real part function. $e$ is the Euler's number. $\mathrm{i}$ is the unit imaginary number. $\langle\cdot,\cdot\rangle$ is the inner product of two vectors. $\overset{p}\longrightarrow$ indicates convergence in probability, while $\overset{d}\longrightarrow$ is convergence in distribution.

\section{Block Sparsity Measures}

\subsection{Definitions}
We firstly introduce some basic concepts for block sparsity and propose a new soft measure of block sparsity.

With $N=\sum_{j=1}^{p}d_j$, we define the $j$-th block $\mathbf{x}[j]$ of a length-$N$ vector $\mathbf{x}$ over $\mathcal{I}=\{d_1,\cdots,d_p\}$. The $j$-th block is of length $d_j$, and the blocks are formed sequentially so that
\begin{align}
\mathbf{x}=(\underbrace{x_1\cdots x_{d_1}}_{\mathbf{x}^{T}[1]}\underbrace{x_{d_1+1}\cdots x_{d_1+d_2}}_{\mathbf{x}^{T}[2]}\cdots\underbrace{x_{N-d_p+1}\cdots x_N}_{\mathbf{x}^{T}[p]})^T. \label{signal}
\end{align}
Without loss of generality, throughout the paper we assume that $d_1=d_2=\cdots=d_p=d$, then $N=pd$. A vector $\mathbf{x}\in\mathbb{R}^N$ is called block $k$-sparse over $\mathcal{I}=\{d,\cdots,d\}$ if $\mathbf{x}[j]$ is nonzero for at most $k$ indices $j$. In other words, by denoting the mixed $\ell_2/\ell_0$ norm
$$
\lVert\mathbf{x}\rVert_{2,0}=\sum_{j=1}^{p}I(\lVert\mathbf{x}[j]\rVert_2>0),
$$
a block $k$-sparse vector $\mathbf{x}$ can be defined by $\lVert\mathbf{x}\rVert_{2,0}\leq k$.

However, in practice this traditional block sparsity measure $\lVert\cdot\rVert_{2,0}$ has a severe drawback of being insensitive to blocks with small entries. For instance, if $\mathbf{x}$ has $k$ blocks with large entries and $p-k$ blocks with small entries, then $\lVert\mathbf{x}\rVert_{2,0}=p$ as soon as they are non-zero. To address this issue, a soft version of the mixed $\ell_2/\ell_0$ norm should be used instead. In this paper we generalize the entropy-based non-block sparsity measure proposed in \cite{l1,l2} to the block sparsity case. 

\begin{definition}
	For any non-zero $\mathbf{x}$ as given in (\ref{signal}) and $\alpha\notin\{0,1,\infty\}$, the soft block sparsity measure is defined in terms of mixed $\ell_2/\ell_\alpha$ norms by \begin{align}
	k_\alpha(\mathbf{x})=\left(\frac{\lVert\mathbf{x}\rVert_{2,\alpha}}{\lVert\mathbf{x}\rVert_{2,1}}\right)^{\frac{\alpha}{1-\alpha}},
	\end{align}
	where the mixed $\ell_2/\ell_\alpha$ norm $\lVert\mathbf{x}\rVert_{2,\alpha}=\left(\sum_{j=1}^{p}\lVert\mathbf{x}[j]\rVert_2^{\alpha}\right)^{1/\alpha}$ for $\alpha>0$. The cases of $\alpha\in\{0,1,\infty\}$ are evaluated as limits: $k_0(\mathbf{x})=\lim\limits_{\alpha\rightarrow 0}k_{\alpha}(\mathbf{x})=\lVert\mathbf{x}\rVert_{2,0}$, $k_1(\mathbf{x})=\lim\limits_{\alpha\rightarrow 1}k_{\alpha}(\mathbf{x})=\exp\left(-\sum_{j=1}^p \frac{\lVert \mathbf{x}[j]\rVert_2}{\lVert \mathbf{x}\rVert_{2,1}}\ln\frac{\lVert \mathbf{x}[j]\rVert_2}{\lVert \mathbf{x}\rVert_{2,1}}\right)$, and $k_\infty(\mathbf{x})=\lim\limits_{\alpha\rightarrow \infty}k_{\alpha}(\mathbf{x})=\frac{\lVert\mathbf{x}\rVert_{2,1}}{\lVert\mathbf{x}\rVert_{2,\infty}}$, where $\lVert\mathbf{x}\rVert_{2,\infty}=\max_{1\leq j\leq p}\lVert\mathbf{x}[j]\rVert_{2}$. 
\end{definition}

In addition, for such a block signal $\mathbf{x}\in\mathbb{R}^N$, a distribution $\pi(\mathbf{x})\in\mathbb{R}^p$ can be induced on the block index set $\{1,\cdots,p\}$, by assigning mass $\pi_j(\mathbf{x})=\lVert\mathbf{x}[j]\rVert_2/\lVert\mathbf{x}\rVert_{2,1}$ at block index $j$ with $\lVert\mathbf{x}\rVert_{2,1}=\sum_{j=1}^{p}\lVert\mathbf{x}[j]\rVert_2$. Hence, we have the following lemma, which establishes the connection between this soft block sparsity measure and the R\'{e}nyi entropy.

\begin{lemma}
	For any $\mathbf{x}\neq \mathbf{0}$ and $\alpha\in[0,\infty]$, its soft block sparsity measure $k_{\alpha}(\mathbf{x})$ can be formulated in terms of entropy by
	\begin{align}
	k_{\alpha}(\mathbf{x})=
	\exp(H_{\alpha}(\pi(\mathbf{x}))),
	\end{align}
	where $H_\alpha$ is the R\'{e}nyi entropy of order $\alpha$ \cite{l2,pv,v}. When $\alpha\notin\{0,1,\infty\}$, the R\'{e}nyi entropy is given explicitly by $H_{\alpha}(\pi(\mathbf{x}))=\frac{1}{1-\alpha}\ln(\sum_{j=1}^p\pi_{j}(\mathbf{x})^\alpha)$, and the cases of $\alpha\in\{0,1,\infty\}$ are defined by evaluating limits, with $H_1$ being the ordinary Shannon entropy $H_{1}(\pi(\mathbf{x}))=-\sum_{j=1}^p \pi_j(\mathbf{x})\ln (\pi_j(\mathbf{x}))$.
\end{lemma}

When the block size $d$ equals 1, our block sparsity measure $k_{\alpha}(\mathbf{x})$ reduces to the non-block sparsity measure $s_\alpha(\mathbf{x})=\left(\lVert\mathbf{x}\rVert_{\alpha}/\lVert\mathbf{x}\rVert_{1}\right)^{\frac{\alpha}{1-\alpha}}$ given by \cite{l2}. Note that the sparsity measure $s_q(\mathbf{x})$ is called $q$-ratio sparsity level of $\mathbf{x}$ in \cite{zhou2018q,zhou2019sparse}, based on which the $q$-ratio constrained minimal singular value (CMSV) is developed for sparse recovery analysis. Very recently, a minimization of $k_q(\cdot)$ (called block $q$-ratio sparsity) has been proposed for block sparse signal recovery in \cite{zhou2021block}.

\begin{figure}[htbp]
	\centering
	\includegraphics[width=\textwidth,height=0.5\textheight]{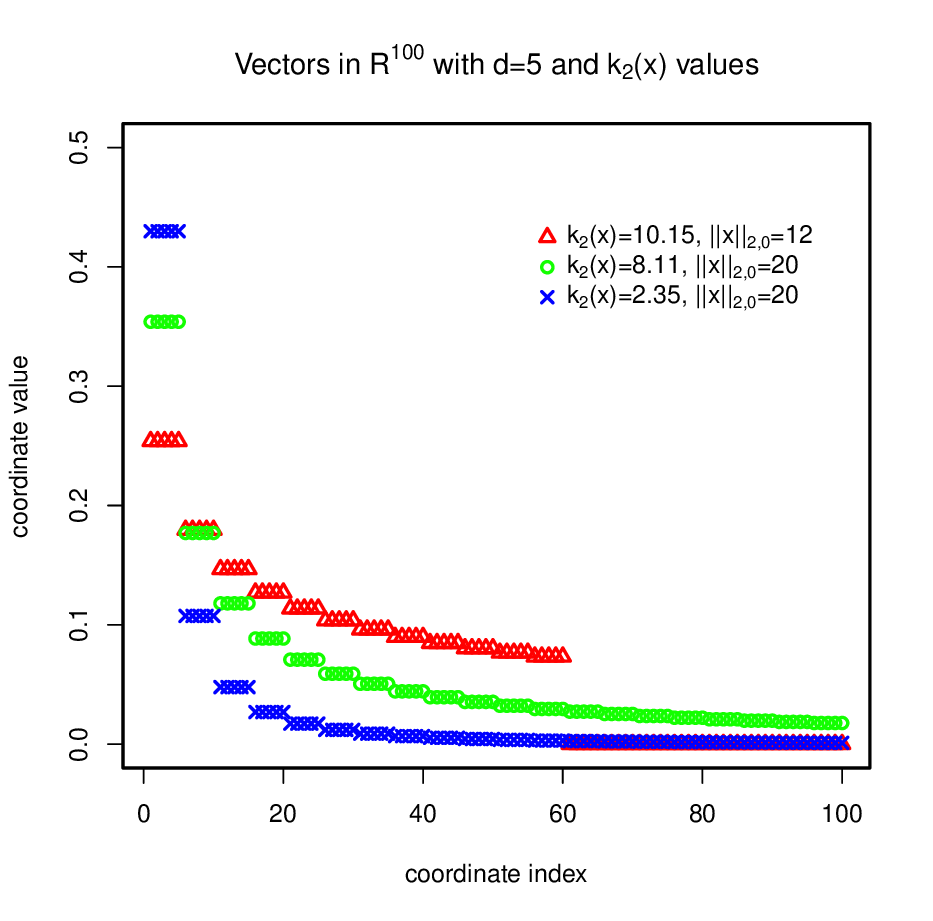}
	\caption{Three vectors (red, green, blue) in $\mathbb{R}^{100}$ are plotted with the $\ell_2$ norm of blocks in decreasing order. We set $d=5$ and compare the values $k_2(\mathbf{x})$ with $\lVert\mathbf{x}\rVert_{2,0}$.}
	\label{fig_1}
\end{figure}

The fact that $k_{\alpha}(\mathbf{x})$ ($\alpha=2$) is a sensible measure of the block sparsity for non-idealized signals is illustrated in Figure \ref{fig_1}. In the case that $\mathbf{x}$ has $k$ blocks with large entries and $p-k$ blocks with small entries, we have $\lVert\mathbf{x}\rVert_{2,0}=p$, whereas $k_{2}(\mathbf{x})\approx k$.

\subsection{Properties}

The block sparsity measure $k_{\alpha}(\mathbf{x})$ has many appealing properties which are similar to those of the non-block sparsity measure $s_{\alpha}(\mathbf{x})$ given in \cite{l2}.
\begin{enumerate}[(a)]
	\item \underline{Continuity}: The function $k_{\alpha}(\cdot)$ is continuous on $\mathbb{R}^N\setminus 0$ for all $\alpha>0$ so that it is stable with respect to small perturbations of the signal.
	\item \underline{Scale-invariance}: It holds that $k_{\alpha}(c\mathbf{x})=k_{\alpha}(\mathbf{x})$ for all $c\neq 0$. As a result, $k_{\alpha}(\mathbf{x})$ relies only on relative (rather than absolute) magnitudes of the entries of $\mathbf{x}$.
	\item \underline{Non-increasing in $\alpha$}: For any $\alpha'\geq\alpha\geq 0$, we have $$
	k_\infty(\mathbf{x})=\frac{\lVert\mathbf{x}\rVert_{2,1}}{\lVert\mathbf{x}\rVert_{2,\infty}}\leq k_{\alpha'}(\mathbf{x})\leq k_{\alpha}(\mathbf{x})\leq k_{0}(\mathbf{x})=\lVert\mathbf{x}\rVert_{2,0},
	$$
	which follows from the non-increasing property of the R\'{e}nyi entropy $H_{\alpha}$ with respect to $\alpha$.
	\item \underline{Range equal to $[1,p]$}: For all $\mathbf{x}\in\mathbb{R}^N\setminus\{0\}$ as given in (\ref{signal}) and all $\alpha\in [0,\infty]$, we have $$
	1\leq k_\infty(\mathbf{x})=\frac{\lVert\mathbf{x}\rVert_{2,1}}{\lVert\mathbf{x}\rVert_{2,\infty}} \leq k_{\alpha}(\mathbf{x})\leq k_{0}(\mathbf{x})=\lVert\mathbf{x}\rVert_{2,0}\leq p.
	$$
\end{enumerate}

\begin{figure}[htbp]
	\centering
	\includegraphics[width=\textwidth,height=0.4\textheight,keepaspectratio]{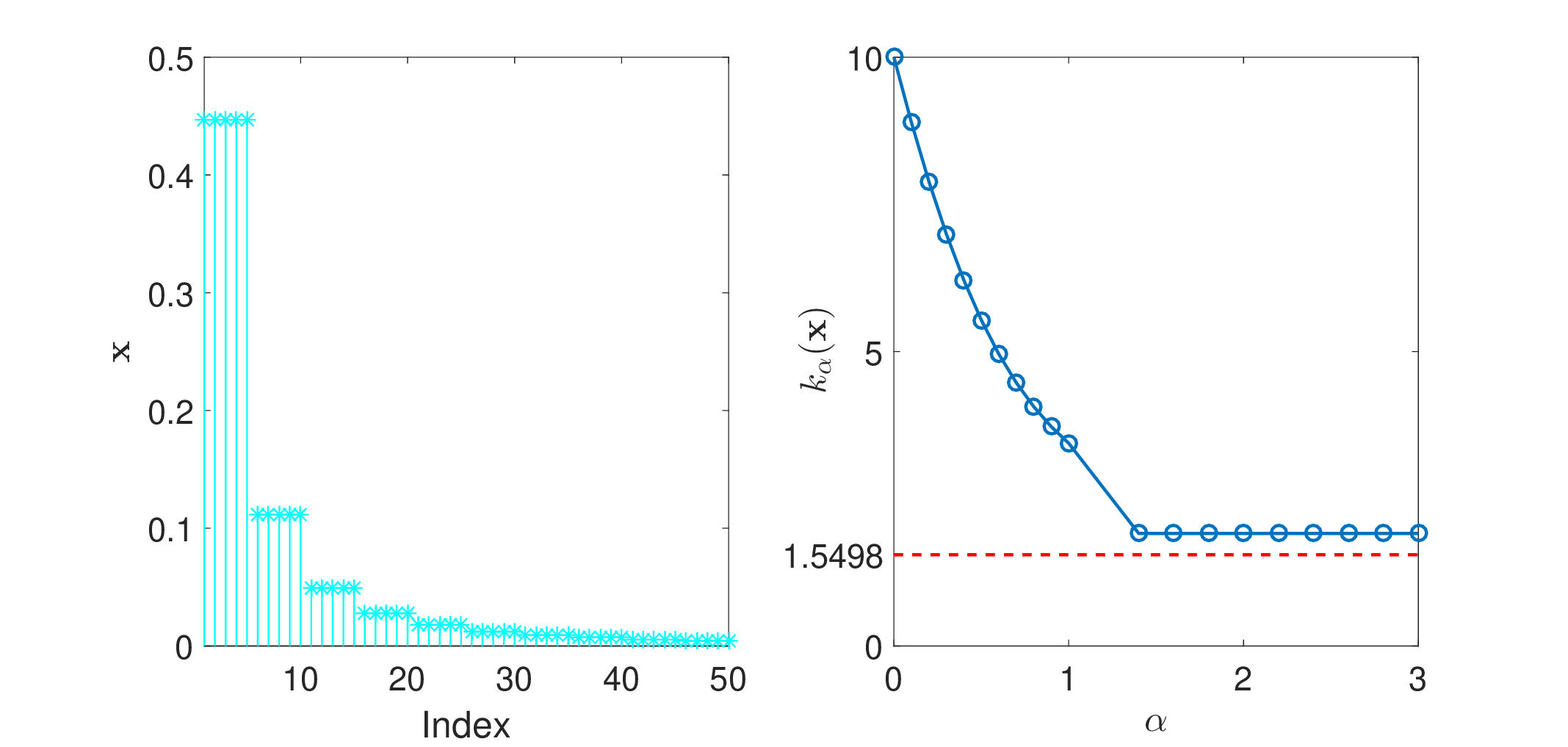}
	\caption{Block sparsity level $k_{\alpha}(\mathbf{x})$ for a block compressible signal $\mathbf{x}$ while varying $\alpha$. Here $\mathbf{x}\in\mathbb{R}^{50}$ is generated with block size being $5$ and entries of each block decay as $i^{-2}/\sqrt{5}$ where $i\in\{1,2,\cdots,10\}$. In the right panel, the dashed red line corresponds to the block sparsity level $k_\infty(\mathbf{x})=\frac{\lVert \mathbf{x}\rVert_{2,1}}{\lVert \mathbf{x}\rVert_{2,\infty}}=1.5498$ }\label{fig:blocksparse}
\end{figure}

To illustrate the properties of this block sparsity measure, we show the block sparsity levels $k_{\alpha}(\cdot)$ for a block compressible signal $\mathbf{x}\in\mathbb{R}^{50}$ displayed in Figure \ref{fig:blocksparse}. As we can see, for a block compressible signal with very small but non-zero entries, $k_{\alpha}(\cdot)$ with a proper $\alpha$ provided a better block sparsity measure compared to the traditional $\lVert \cdot\rVert_{2,0}$. In addition, the block sparsity level $k_{\alpha}(\mathbf{x})$ is non-increasing with respect to $\alpha$ and its value range is between $k_{\infty}(\mathbf{x})=1.5498$ and $k_0(\mathbf{x})=\lVert \mathbf{x}\rVert_{2,0}=10$.

\subsection{An Error Bound in terms of $k_{2}(\cdot)$}

Before presenting the estimation procedure for the $\lVert\mathbf{x}\rVert_{2,\alpha}^{\alpha}$ and $k_{\alpha}(\mathbf{x})$ with $\alpha\in(0,2]$, we give the block sparse signal recovery results in terms of $k_2(\mathbf{x})$. To recover the block sparse signal in CS model (\ref{1.1}), here we use the following mixed $\ell_2/\ell_1$ norm optimization algorithm proposed in \cite{ekb,em}: \begin{align}
\widehat{\mathbf{x}}=\mathop{\arg\min}_{\mathbf{x}\in\mathbb{R}^N}\,\lVert\mathbf{x}\rVert_{2,1},\,\,\,\text{subject to}\,\,\,\lVert\mathbf{y}-A\mathbf{x}\rVert_2\leq\delta,
\label{2.1}
\end{align}
where $\mathbf{y}=A\mathbf{x}+\boldsymbol{\varepsilon}$ and $\delta\geq 0$ is an upper bound on the noise level $\lVert\boldsymbol{\varepsilon}\rVert_2$. Then, we have the following result concerning the robust recovery for block sparse signals. 

\begin{lemma}(\cite{em}) 
	Let $\mathbf{y}=A\mathbf{x}+\boldsymbol{\varepsilon}$ be noisy measurements of a vector $\mathbf{x}$ and fix a number $k\in\{1,\cdots,p\}$. Let $\mathbf{x}^k$ denote the best block $k$-sparse approximation of $\mathbf{x}$, such that $\mathbf{x}^k$ is block $k$-sparse and minimizes $\lVert\mathbf{x}-\mathbf{f}\rVert_{2,1}$ over all the block $k$-sparse vectors $\mathbf{f}$, and let $\widehat{\mathbf{x}}$ be a solution to (\ref{2.1}), a Gaussian random matrix $A$ of size $m\times N$ with entries $A_{ij}\sim N(0,\frac{1}{m})$, and block sparse signals over $\mathcal{I}=\{d_1=d,\cdots,d_p=d\}$, where $N=pd$ for some integer $p$. Then, there are constants $c_0,c_1,c_2,c_3>0$, such that with probability at least $1-2\exp(-c_1 m)$,  we have\begin{align}
	\frac{\lVert\widehat{\mathbf{x}}-\mathbf{x}\rVert_2}{\lVert\mathbf{x}\rVert_2}\leq c_2\frac{\lVert\mathbf{x}-\mathbf{x}^k\rVert_{2,1}}{\sqrt{k}\lVert\mathbf{x}\rVert_2}+c_3\frac{\delta}{\lVert\mathbf{x}\rVert_2},\label{2.2}
	\end{align}
	whenever $m\geq c_0 k\ln(eN/kd)$.
\end{lemma}

\noindent\\
{\bf Remark 1.} Note that the first term on the right-hand side of (\ref{2.2}) results from the fact that $\mathbf{x}$ is not exactly block $k$-sparse, while the second term quantifies the recovery error due to the measurement noise. When the block size $d=1$, this lemma goes to the conventional CS result for non-block sparse signal recovery. Explicit use of block sparsity reduces the required number of measurements from $\mathrm{O}(kd\ln(eN/kd))$ to $\mathrm{O}(k\ln(eN/kd))$ by $d$ times.\\

However, the relative error bound established above has the drawback that the ratio term $\frac{\lVert\mathbf{x}-\mathbf{x}^k\rVert_{2,1}}{\sqrt{k}\lVert\mathbf{x}\rVert_2}$ is often unknown. As a result, it is unclear how large $m$ should be chosen to make the relative error small. In order to address this issue, in the following lemma we present an upper bound for the relative error in terms of $m$ and the new estimable block sparsity measure $k_2(\mathbf{x})$.

\begin{lemma}
	Let $\mathbf{y}=A\mathbf{x}+\boldsymbol{\varepsilon}$ be noisy measurements of a vector $\mathbf{x}$, and let $\widehat{\mathbf{x}}$ be a solution to (\ref{2.1}), a Gaussian random matrix $A$ of size $m\times N$ with entries $A_{ij}\sim N(0,\frac{1}{m})$, and block sparse signals over $\mathcal{I}=\{d_1=d,\cdots,d_p=d\}$, where $N=pd$ for some integer $p$. Then, there are constants $\tau_0,\tau_1,\tau_2,\tau_3>0$ such that with probability at least $1-2\exp(-\tau_1 m)$, we have \begin{align}
	\frac{\lVert\widehat{\mathbf{x}}-\mathbf{x}\rVert_2}{\lVert\mathbf{x}\rVert_2}\leq \tau_2\sqrt{\frac{k_2(\mathbf{x})d\ln(\frac{eN}{m})}{m}}+\tau_3\frac{\delta}{\lVert\mathbf{x}\rVert_2},
	\end{align}
	whenever $m$ and $N$ satisfy $\tau_0\ln(\tau_0\frac{eN}{m})\leq m\leq N$.
\end{lemma}
\bigskip

\section{Block Sparsity Estimation}

Let us now turn to the estimation of block sparsity $k_{\alpha}(\cdot)$ with $\alpha\in(0,2]$. There are two reasons to consider this interval. One is that small $\alpha$ is usually a better block sparsity measure than very large $\alpha$ in applications. And we can approximate $\lVert\cdot\rVert_{2,0}$ by $k_{\alpha}(\cdot)$ with very small $\alpha$ as will be shown later. The other reason is that our estimation method relies on the $\alpha$-stable distribution, which requires $\alpha$ to lie in $(0,2]$. The core idea to obtain the block sparsity estimators is using random projections. In contrast to the conventional non-block sparsity estimation by using projections with univariate symmetric $\alpha$-stable random variables \cite{l2,z}, we use projections with multivariate centered isotropic symmetric $\alpha$-stable random vectors for the block sparsity estimation.

\subsection{Multivariate Isotropic Stable Distribution}

We start with the definition of multivariate centered isotropic symmetric $\alpha$-stable distribution.

\begin{figure}[!t]
	\centering
	\includegraphics[width=\textwidth,height=0.4\textheight]{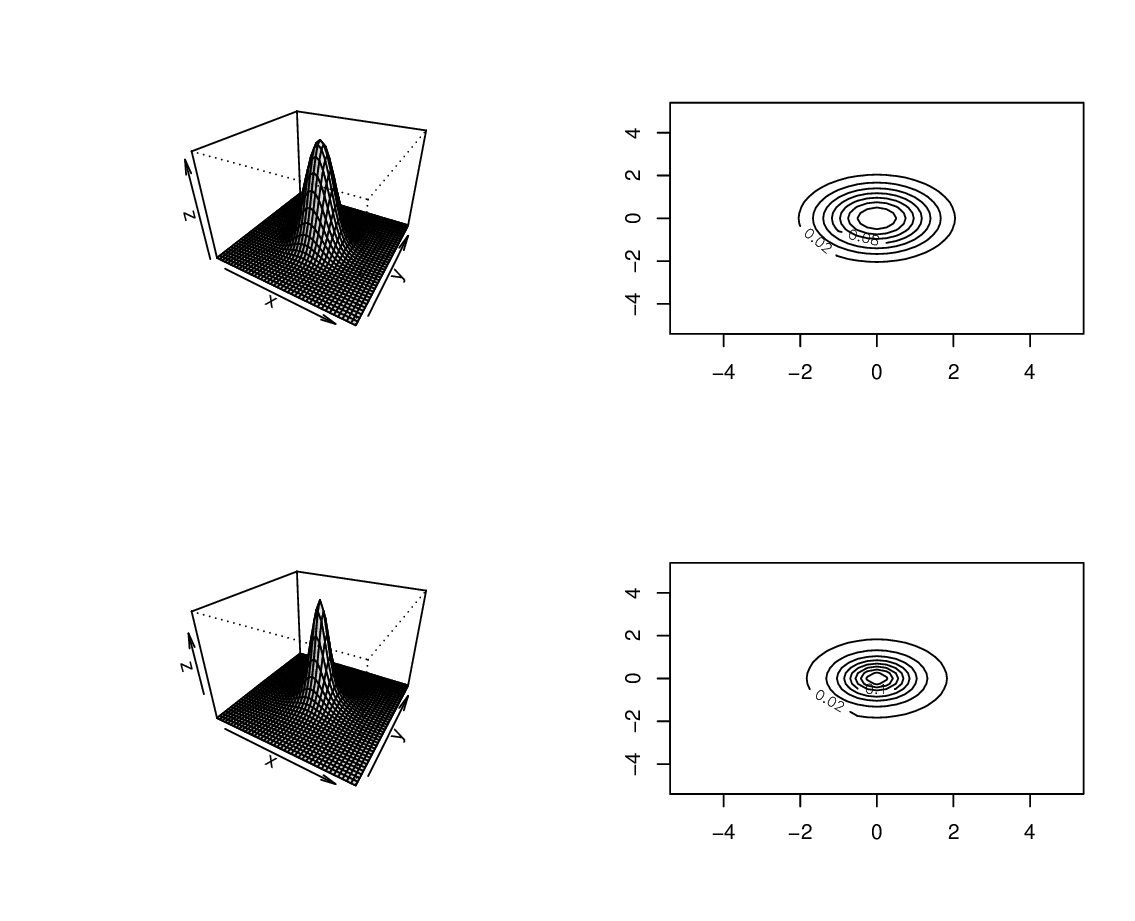}
	\caption{Perspective and Contour Plots for the Bivariate Centered Isotropic Symmetric Stable Densities. The top ones are for the Bivariate Independent Gaussian Distribution, while the bottom ones are for the Bivariate Spherical Symmetric Cauchy Distribution.}
	\label{fig_2}
\end{figure}

\begin{definition} (\cite{n,p})
	For $d\geq 1$, a $d$-dimensional random vector $\mathbf{v}$ has a multivariate centered isotropic symmetric $\alpha$-stable distribution if there are constants $\gamma>0$ and $\alpha\in(0,2]$ such that its characteristic function has the form \begin{align}
	E[\exp(\mathrm{i}\mathbf{u}^{T}\mathbf{v})]=\exp(-\gamma^{\alpha}\lVert\mathbf{u}\rVert_2^\alpha),\,\,\,\text{for all $\mathbf{u}\in\mathbb{R}^d$}.
	\end{align}
	We denote the distribution by $\mathbf{v}\sim S(d,\alpha,\gamma)$, and $\gamma$ is referred to as the scale parameter.
\end{definition}

\noindent\\
{\bf Remark 2.} The most well-known example of multivariate centered isotropic symmetric stable distribution is the case of $\alpha=2$ (Multivariate Independent Gaussian Distribution), and in this case, the components of the Multivariate Gaussian random vector are independent. Another case is $\alpha=1$ (Multivariate Spherical Symmetric Cauchy Distribution \cite{p}), unlike the Multivariate Independent Gaussian case, the components of Multivariate Spherical Symmetric Cauchy random vector are uncorrelated, but dependent. The perspective and contour plots for the densities of these two cases are illustrated in Figure \ref{fig_2}. The multivariate centered isotropic symmetric $\alpha$-stable random vector is a direct extension of the univariate symmetric $\alpha$-stable random variable, which is the special case when the dimension parameter $d=1$. In practice, to simulate a $d$-dimensional random vector $\mathbf{v}$ from the multivariate centered isotropic symmetric $\alpha$-stable distribution $S(d,\alpha,\gamma)$, we can adopt the fact that $\mathbf{v}=D^{1/2}\mathbf{q}$, where $D\sim \tilde{S}(1,\alpha/2,2\gamma^2[\cos(\pi\alpha/4)]^{2/\alpha})$ is an independent univariate positive
$(\alpha/2)$-stable random variable and $\mathbf{q}\sim N(0,\mathrm{I}_d)$ is a standard $d$-dimensional Gaussian random vector, see \cite{n} for more details.

\subsection{Estimation Procedure}

The most important finding in our paper is that by adopting random projections using i.i.d. multivariate centered isotropic symmetric $\alpha$-stable random vectors, we can directly generalize the estimation procedure for the non-block sparsity measure given in \cite{l2} and obtain the estimators for $\lVert\cdot\rVert_{2,\alpha}^{\alpha}$ and $k_{\alpha}(\cdot)$ with $\alpha\in(0,2]$. In the subsection that follows, we will present the corresponding estimation procedure.

We estimate the $\lVert\mathbf{x}\rVert_{2,\alpha}^{\alpha}$ based on the random linear projection measurements: \begin{align}
y_i=\langle \mathbf{a}_i,\mathbf{x}\rangle+\sigma\varepsilon_i, \,\,\,i=1,2,\cdots,n, \label{measurements}
\end{align}
where $\mathbf{a}_i\in\mathbb{R}^N$ is an i.i.d. random vector, and $\mathbf{a}_i=(\mathbf{a}_{i1}^T,\cdots,\mathbf{a}_{ip}^T)^T$ with $\mathbf{a}_{ij},j\in\{1,\cdots,p\}$ i.i.d. drawn from $S(d,\alpha,\gamma)$. The noise terms $\{\varepsilon_i, i=1,\cdots,n\}$ are i.i.d. generated from a distribution $F_0$ with its characteristic function being $\varphi_0$. The variable sets $\{\varepsilon_1,\cdots,\varepsilon_n\}$ and $\{\mathbf{a}_1,\cdots,\mathbf{a}_n\}$ are independent. In addition, we assume that $\{\varepsilon_i, i=1,2,\cdots,n\}$ are symmetric about $0$, with $0<E|\varepsilon_1|<\infty$, but they may have infinite variance. We impose a minor technical constraint on $F_0$ that the roots of its characteristic function $\varphi_0$ are isolated (i.e. no limit points). It is easy to verify that many families of distributions, including Gaussian, Student's $t$, Laplace, uniform$[a, b]$, and stable
laws, satisfy this condition. For simplicity, throughout this paper we assume that the noise scale parameter $\sigma\geq0$ and the distribution $F_0$ are known .

Because our work involves multiple values of $\alpha$, next we will use $\gamma_\alpha$ instead of $\gamma$ to avoid confusion. Before proceeding to the detailed estimation procedure for $\lVert\cdot\rVert_{2,\alpha}^{\alpha}$ and $k_{\alpha}(\cdot)$, it is necessary to obtain the following key lemma that establishes the relationship between the multivariate isotropic $\alpha$-stable distribution and the mixed $\ell_2/\ell_{\alpha}$ norm $\lVert\cdot\rVert_{2,\alpha}$.

\begin{lemma}
	Let $\mathbf{x}=(\mathbf{x}[1]^T,\cdots,\mathbf{x}[p]^T)^T\in\mathbb{R}^N$ be fixed, and suppose $\mathbf{a}_1=(\mathbf{a}_{11}^T,\cdots,\mathbf{a}_{1p}^T)^T$ with $\mathbf{a}_{1j}, j\in\{1,\cdots,p\}$ i.i.d. drawn from $S(d,\alpha,\gamma_\alpha)$ with $\alpha\in(0,2]$
	and $\gamma_\alpha>0$. Then, the random variable $\langle \mathbf{a}_1,\mathbf{x}\rangle$ has the distribution $S(1,\alpha,\gamma_\alpha\lVert\mathbf{x}\rVert_{2,\alpha})$.
\end{lemma}

\noindent\\
{\bf Remark 3.} If $\mathbf{x}=(\mathbf{x}[1]^T,\cdots,\mathbf{x}[p]^T)^T\in\mathbb{R}^N$ has different block lengths which are $\{d_1,d_2,\cdots,d_p\}$ respectively, then we need to choose the projection random vector $\mathbf{a}_1=(\mathbf{a}_{11}^T,\cdots,\mathbf{a}_{1p}^T)^T$ with $\mathbf{a}_{1j}, j\in\{1,\cdots,p\}$ i.i.d. drawn from $S(d_j,\alpha,\gamma_\alpha)$. In that case, the conclusion in this lemma and all the results in what follows still hold without any modifications. 

\noindent\\
{\bf Remark 4.} Lemma 1 in \cite{l2} is a special case of this lemma when the block size is one. \\

From this lemma, we can get that if $\tilde{y}_i=\langle\mathbf{a}_i,\mathbf{x}\rangle$ with a set of i.i.d. measurement random vectors $\{\mathbf{a}_1,\cdots,\mathbf{a}_n\}$ given as mentioned above, then $\{\tilde{y}_1,\cdots,\tilde{y}_n\}$ is an i.i.d. sample from the distribution $S(1,\alpha,\gamma_\alpha\lVert\mathbf{x}\rVert_{2,\alpha})$. Therefore, the problem of estimating the norm $\lVert\mathbf{x}\rVert_{2,\alpha}^{\alpha}$ from noiseless random linear measurements reduces to a problem of estimating the scale parameter of a univariate stable distribution from an i.i.d. sample.

Now we are ready to extend this idea and present the estimation procedure for $\lVert\cdot\rVert_{2,\alpha}^{\alpha}$ and $k_{\alpha}(\cdot)$ by adopting the characteristic function method \cite{l2,mh,mhl}. We use two separate sets of measurements to estimate $\lVert\mathbf{x}\rVert_{2,1}$ and
$\lVert\mathbf{x}\rVert_{2,\alpha}^{\alpha}$ with their corresponding estimators denoted as $\widehat{\lVert\mathbf{x}\rVert_{2,1}}$ and
$\widehat{\lVert\mathbf{x}\rVert_{2,\alpha}^{\alpha}}$. We denote the sample sizes of these two measurements by $n_1$ and $n_{\alpha}$, respectively. In order to simplify the discussion, we only introduce the procedure to obtain $\widehat{\lVert\mathbf{x}\rVert_{2,\alpha}^{\alpha}}$ for any $\alpha\in(0,2]$ with $\alpha=1$ being a special case. We suggest the following estimator for $k_\alpha(\mathbf{x})$ with $\alpha\in(0,2]\setminus\{1\}$ by integrating these two estimators $\widehat{\lVert\mathbf{x}\rVert_{2,1}}$ and $\widehat{\lVert\mathbf{x}\rVert_{2,\alpha}^{\alpha}}$:
\begin{align}
\hat{k}_{\alpha}(\mathbf{x})=\frac{\left(\widehat{\lVert\mathbf{x}\rVert_{2,\alpha}^{\alpha}}\right)^{\frac{1}{1-\alpha}}}
{\left(\widehat{\lVert\mathbf{x}\rVert_{2,1}}\right)^{\frac{\alpha}{1-\alpha}}}.
\end{align}

With regard to the estimation of $\lVert\mathbf{x}\rVert_{2,\alpha}^{\alpha}$ based on the noisy random linear projection measurements (\ref{measurements}), the characteristic function of $\{y_i,i=1,\cdots,n_{\alpha}\}$ has the form: \begin{align}
\Psi(t)=E[\exp(\mathrm{i}ty_i)]&=E[\exp(\mathrm{i}\langle a_i, \mathbf{x}\rangle)]\cdot E[\exp(\mathrm{i}t\sigma \varepsilon_i)] \nonumber \\
&=\exp(-\gamma_\alpha^{\alpha}\lVert\mathbf{x}\rVert_{2,\alpha}^{\alpha}|t|^{\alpha})\cdot\varphi_0(\sigma t),
\end{align}
where $t\in\mathbb{R}$. Then, we have the estimation equation
$$\lVert\mathbf{x}\rVert_{2,\alpha}^{\alpha}=-\frac{1}{\gamma_\alpha^{\alpha}|t|^\alpha}\ln \left|\mathrm{Re}\left(\frac{\Psi(t)}{\varphi_0(\sigma t)}\right)\right|.
$$
By adopting the empirical characteristic function $$\hat{\Psi}_{n_\alpha}(t)=\frac{1}{n_\alpha}\sum\limits_{i=1}^{n_\alpha}e^{\mathrm{i}ty_i}$$
to estimate $\Psi(t)$, we are able to obtain the estimator of $\lVert\mathbf{x}\rVert_{2,\alpha}^{\alpha}$ given by \begin{align}
\widehat{\lVert\mathbf{x}\rVert_{2,\alpha}^{\alpha}}=:\hat{v}_{\alpha}(t)=-\frac{1}{\gamma_\alpha^{\alpha}|t|^\alpha}\ln \left|\mathrm{Re}\left(\frac{\hat{\Psi}_{n_\alpha}(t)}{\varphi_0(\sigma t)}\right)\right|,
\end{align}
when $t\neq 0$ and $\varphi_0(\sigma t)\neq 0$. 

\subsection{Asymptotic Properties}

This subsection draws together the large sample properties for the proposed estimators. Let the noise-to-signal ratio constant be $\rho_\alpha=\frac{\sigma}{\gamma_\alpha\lVert\mathbf{x}\rVert_{2,\alpha}}$. We have the following uniform central limit theorem (CLT) \cite{du} for $\hat{v}_\alpha(t)$.

\begin{theorem} 
	Let $\alpha\in(0,2]$ and $\hat{t}$ be any function of $\{y_1,\cdots,y_{n_\alpha}\}$ such that \begin{align}
	\gamma_\alpha\hat{t}\lVert \mathbf{x}\rVert_{2,\alpha}\overset{p}\longrightarrow c_\alpha,
	\end{align}
	as $(n_\alpha,N)\rightarrow \infty$ for some finite constant $c_\alpha\neq 0$ and $\varphi_0(\rho_\alpha c_\alpha)\neq 0$. Then, it holds that \begin{align}
	\sqrt{n_\alpha}\left(\frac{\hat{v}_{\alpha}(\hat{t})}{\lVert\mathbf{x}\rVert_{2,\alpha}^{\alpha}}-1\right)\overset{d}\longrightarrow N(0,\theta_\alpha(c_\alpha,\rho_\alpha))
	\end{align}
	as $(n_{\alpha},N)\rightarrow\infty$, where the limiting variance $\theta_\alpha(c_\alpha,\rho_\alpha)$ is strictly positive and defined by \begin{align}
	\theta_\alpha(c_\alpha,\rho_\alpha)=\frac{1}{|c_\alpha|^{2\alpha}}\Big(\frac{\exp(2|c_\alpha|^{\alpha})}{2\varphi_0(\rho_\alpha
		|c_\alpha|)^2} +\frac{\varphi_0(2\rho_\alpha|c_\alpha|)}{2\varphi_0(\rho_\alpha|c_\alpha|)^2}\exp((2-2^\alpha)|c_\alpha|^\alpha)-1\Big).
	\end{align}
\end{theorem}

Because it is easy to implement and produces a fairly good estimator, throughout this paper we use  $\hat{t}_{\mathrm{pilot}}=\min\{\frac{1}{\hat{m}_\alpha},\frac{\eta_0}{\sigma}\}$ as our $\hat{t}$, where $\eta_0>0$ is any number such that $\varphi_0(\eta)>\frac{1}{2}$ for all $\eta\in[0,\eta_0]$ and  $\hat{m}_\alpha=\mathrm{median}\{|y_1|,\cdots,|y_{n_{\alpha}}|\}$ is the median absolute deviation statistic. Then we can obtain the consistent estimator $\hat{c}_\alpha=\gamma_\alpha\hat{t}_{\mathrm{pilot}}[\hat{v}_{\alpha}(\hat{t}_{\mathrm{pilot}})]^{1/\alpha}$ of a constant $c_\alpha=\min\left(\frac{1}{\mathrm{median}(|S_1+\rho_\alpha\varepsilon_1|)},\frac{\eta_0}{\rho_\alpha}\right)$, where random variable $S_1\sim S(1,1,1)$ (see the Proposition 3 in \cite{l2}), and the consistent estimator of $\rho_\alpha$, $\hat{\rho}_\alpha=\frac{\sigma}{\gamma_\alpha[\hat{v}_{\alpha}(\hat{t}_{\mathrm{pilot}})]^{1/\alpha}}$. Therefore, the consistent estimator of the limiting variance $\theta_\alpha(c_\alpha,\rho_\alpha)$ is $\theta_\alpha(\hat{c}_\alpha,\hat{\rho}_\alpha)$. 

According to these results, we can derive the following corollary, in which a confidence interval for $\lVert\mathbf{x}\rVert_{2,\alpha}^{\alpha}$ is constructed.

\begin{cor}
	Under the conditions of Theorem 1, as $(n_\alpha, N)\rightarrow \infty$, we have \begin{align}
	\sqrt{\frac{n_\alpha}{\theta_\alpha(\hat{c}_\alpha,\hat{\rho}_\alpha)}}\left(\frac{\hat{v}_{\alpha}(\hat{t}_{\mathrm{pilot}})}{\lVert\mathbf{x}\rVert_{2,\alpha}^{\alpha}}-1\right)\overset{d}\longrightarrow N(0,1). \label{clt}
	\end{align}
	Consequently, the asymptotic $1-\beta$ confidence interval for $\lVert\mathbf{x}\rVert_{2,\alpha}^{\alpha}$ is \begin{align}
	\bigg[\Big(1-\sqrt{\frac{\theta_\alpha(\hat{c}_\alpha,\hat{\rho}_\alpha)}{n_\alpha}}z_{1-\beta/2}\Big)\hat{v}_{\alpha}(\hat{t}_{\mathrm{pilot}}), 
	\Big(1+\sqrt{\frac{\theta_\alpha(\hat{c}_\alpha,\hat{\rho}_\alpha)}{n_\alpha}}z_{1-\beta/2}\Big)\hat{v}_{\alpha}(\hat{t}_{\mathrm{pilot}})\bigg], \label{conf_interval}
	\end{align}
	where $z_{1-\beta/2}$ is the $(1-\beta/2)$-quantile of the standard normal distribution.
\end{cor}

It should be noted that the simple confidence interval given above is obtained using a CLT for the reciprocal $\frac{\lVert\mathbf{x}\rVert_{2,\alpha}^{\alpha}}{\hat{v}_{\alpha}(\hat{t}_{\mathrm{pilot}})}$ via the delta method. Following this corollary, we are ready to obtain the asymptotic results for $\hat{k}_\alpha(\mathbf{x})$ with $\alpha\in(0,2]\setminus\{1\}$, by combining the estimators $\hat{v}_{\alpha}$ and $\hat{v}_1$ with their corresponding $\hat{t}_{\mathrm{pilot}}$ values. For each $\alpha\in(0,2]\setminus\{1\}$, we make the assumption that there is a constant $\bar{\pi}_\alpha\in (0,1)$, such that as $(n_1,n_\alpha,N)\rightarrow\infty$, $$
\pi_\alpha:=\frac{n_\alpha}{n_1+n_\alpha}=\bar{\pi}_\alpha+o(n_\alpha^{-1/2}).
$$

\begin{theorem}
	Let $\alpha\in(0,2]\setminus\{1\}$ and assume that the conditions of Theorem 1 hold. Then as $(n_1,n_\alpha,N)\rightarrow\infty$, we have \begin{align}
	\sqrt{\frac{n_1+n_\alpha}{\hat{w}_\alpha}}\left(\frac{\hat{k}_\alpha(\mathbf{x})}{k_{\alpha}(\mathbf{x})}-1\right)\overset{d}\longrightarrow N(0,1),
	\end{align}
	where $\hat{w}_\alpha=\frac{\theta_\alpha(\hat{c}_\alpha,\hat{\rho}_{\alpha})}{\pi_\alpha}(\frac{1}{1-\alpha})^2+\frac{\theta_1(\hat{c}_1,\hat{\rho}_1)}{1-\pi_\alpha}(\frac{\alpha}{1-\alpha})^2$. Consequently, the asymptotic $1-\beta$ confidence interval for $k_{\alpha}(\mathbf{x})$ is \begin{align}
	\bigg[\Big(1-\sqrt{\frac{\hat{w}_\alpha}{n_1+n_\alpha}}z_{1-\beta/2}\Big)\hat{k}_\alpha(\mathbf{x}),
	\Big(1+\sqrt{\frac{\hat{w}_\alpha}{n_1+n_\alpha}}z_{1-\beta/2}\Big)\hat{k}_\alpha(\mathbf{x})\bigg],
	\end{align}
	where $z_{1-\beta/2}$ is the $(1-\beta/2)$-quantile of the standard normal distribution. \\
\end{theorem}

\subsection{Estimating $\lVert\cdot\rVert_{2,0}$ by with Small $\alpha$}

This subsection presents the approximation of $\hat{k}_\alpha(\mathbf{x})$ to $\lVert\mathbf{x}\rVert_{2,0}$ when $\alpha$ is close to $0$. We begin with defining the block dynamic range (BDNR) of a non-zero signal $\mathbf{x}\in\mathbb{R}^N$ (as given in (\ref{signal})) by \begin{align}
\mathrm{BDNR}(\mathbf{x})=\frac{\lVert\mathbf{x}\rVert_{2,\infty}}{|\mathbf{x}|_{2,\min}},
\end{align}
where $|\mathbf{x}|_{2,\min}=\min\{\lVert\mathbf{x}[j]\rVert_2:\mathbf{x}[j]\neq \mathbf{0}, j=1,\cdots,p\}$ denotes the smallest $\ell_2$ norm of the non-zero block of $\mathbf{x}$. Note that our $\mathrm{BDNR}(\mathbf{x})$ includes the dynamic range (DNR) defined in \cite{l2} as a special case of block size $d=1$. Then we obtain the approximation result as follows.

\begin{theorem}
	Let $\alpha\in(0,1)$, $\mathbf{x}\in\mathbb{R}^N$ is non-zero signal as given in (\ref{signal}), and let $\tilde{k}_{\alpha}(\mathbf{x})$ be any real number. Then, it holds that \begin{align}
	\left|\frac{\tilde{k}_{\alpha}(\mathbf{x})}{\lVert\mathbf{x}\rVert_{2,0}}-1\right|\leq \left|\frac{\tilde{k}_{\alpha}(\mathbf{x})}{k_{\alpha}(\mathbf{x})}-1\right|
	+\frac{\alpha}{1-\alpha}\Big(\ln(\mathrm{BDNR}(\mathbf{x}))+\alpha\ln(\lVert\mathbf{x}\rVert_{2,0})\Big). \label{th3}
	\end{align}
\end{theorem}

\noindent\\
{\bf Remark 5.} Note that the approximation result established above involves no randomness and it applies to any estimator $\tilde{k}_{\alpha}(\mathbf{x})$. The first term on the right-hand side of (\ref{th3}) can be controlled by Theorem 2, when $\tilde{k}_{\alpha}(\mathbf{x})$ is chosen as our proposed estimator $\hat{k}_{\alpha}(\mathbf{x})$. The second term on the right-hand side of (\ref{th3}) is the approximation error, which decreases as $\alpha$ decreases. In addition, when the $\ell_2$ norms of the signal's blocks are similar, the quantity of $\ln(\mathrm{BDNR}(\mathbf{x}))$ will not be too large so that the bound behaves well and estimating $\lVert\mathbf{x}\rVert_{2,0}$ is of interest. Whereas, if there is a big difference in the $\ell_2$ norms of the signal's blocks, then $\ln(\mathrm{BDNR}(\mathbf{x}))$ will be very large such that $\lVert\mathbf{x}\rVert_{2,0}$ may not be the best block sparsity measure to estimate.

\section{Simulations}
In this section, we conduct numerical experiments to illustrate our theoretical results. We start with the discussion of $\alpha=2$, that is, we use $\hat{k}_2(\mathbf{x})$ to estimate the block sparsity measure $k_2(\mathbf{x})$. To obtain the estimator $\hat{k}_2(\mathbf{x})$, we require a set of $n_1$ measurements by using the multivariate centered isotropic symmetric Cauchy projection, and a set of $n_2$ measurements by using the multivariate centered isotropic symmetric Gaussian projection. The samples $\mathbf{y}_1\in\mathbb{R}^{n_1}$ and $\mathbf{y}_2\in\mathbb{R}^{n_2}$ are generated as:
\begin{align}
\mathbf{y}_1=A_1\mathbf{x}+\sigma\boldsymbol{\varepsilon}_1\,\,\,\text{and}\,\,\,\mathbf{y}_2=A_2\mathbf{x}+\sigma\boldsymbol{\varepsilon}_2,
\end{align}
where $A_1=(\mathbf{a}_1,\cdots,\mathbf{a}_{n_1})\in\mathbb{R}^{n_1\times N}$ with $\mathbf{a}_i\in\mathbb{R}^N$ being i.i.d. random vector, and $\mathbf{a}_i=(\mathbf{a}_{i1}^T,\cdots,\mathbf{a}_{ip}^T)^T$ with $\mathbf{a}_{ij}, j\in\{1,\cdots,p\}$ i.i.d. drawn from $S(d,1,\gamma_1)$ with $\gamma_1=1$. Similarly,  $A_2=(\mathbf{b}_1,\cdots,\mathbf{b}_{n_2})\in\mathbb{R}^{n_2\times N}$ with $\mathbf{b}_i\in\mathbb{R}^N$ being i.i.d. random vector, and $\mathbf{b}_i=(\mathbf{b}_{i1}^T,\cdots,\mathbf{b}_{ip}^T)^T$ with $\mathbf{b}_{ij}, j\in\{1,\cdots,p\}$ i.i.d. drawn from $S(d,2,\gamma_2)$ with $\gamma_2=\frac{\sqrt{2}}{2}$. The noise terms $\boldsymbol{\varepsilon}_1$ and $\boldsymbol{\varepsilon}_2$ are generated with i.i.d entries from a standard normal distribution. In this set of experiments, we consider a sequence of pairs for the sample sizes $(n_1,n_2)=(50,50),(100,100),(200,200),\cdots,(500,500)$. For each experiment, we replicated $200$ times. Consequently, we have $200$ realizations of $\hat{k}_2(\mathbf{x})$ for each $(n_1,n_2)$. Throughout all the experiments, we fix $\eta_0=1$. 

\subsection{Exactly Block Sparse Case}
\begin{figure}[!t]
	\centering
	\includegraphics[width=\textwidth,height=0.7\textheight,keepaspectratio]{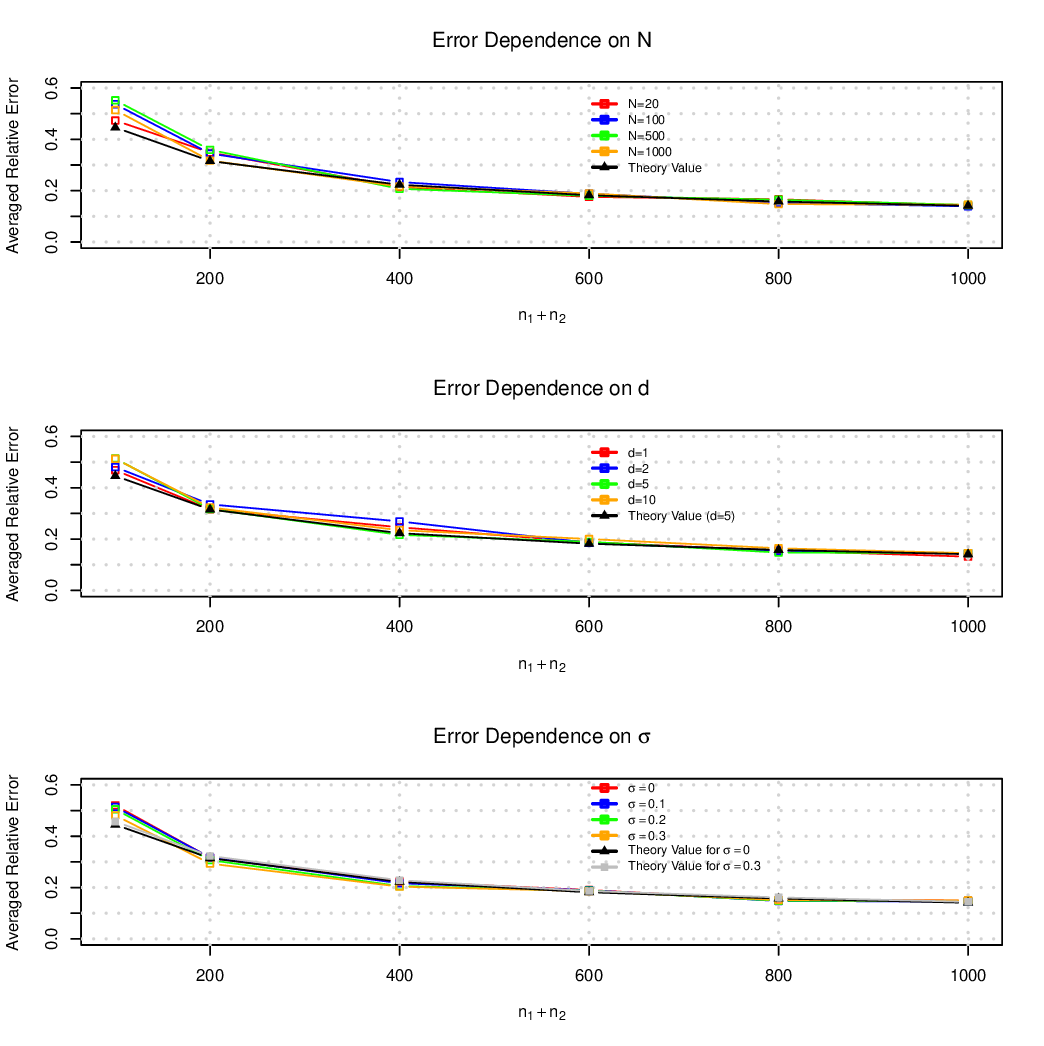}
	\caption{The averaged relative error $|\frac{\hat{k}_2(\mathbf{x})}{k_2(\mathbf{x})}-1|$ depending on $N$, $d$ and $\sigma$ for the exactly block sparse case.}
	\label{fig_3}
\end{figure}

First, we let our signal $\mathbf{x}$ be the following simple exactly block sparse vector
$$
\mathbf{x}=(\frac{1}{\sqrt{10}}\mathbf{1}_{10}^T, \mathbf{0}_{N-10}^T)^T,
$$
where $\mathbf{1}_q$ is a vector of length $q$ with entries all ones, $\mathbf{0}_q$ is the zero vector of length $q$. Then it is obvious that $\lVert\mathbf{x}\rVert_{2,2}=\lVert\mathbf{x}\rVert_{2}=1$, while $\lVert\mathbf{x}\rVert_{2,1}$ and $k_2(\mathbf{x})$ depend on the block size $d$ that we choose. Under this exactly block sparse setting, we have performed the following two simulation studies on the error dependence on the parameters and the asymptotic normality of the proposed estimators.

a) The first simulation study in this part is conducted to illustrate the influence of varying the parameter settings. It is done with a variety of parameters, like $N$, $d$ and $\sigma$, each of which corresponds to a separate plot in Figure \ref{fig_3}. Except in the top plot where $N=20,100,500,1000$, the signal dimension $N$ is set to $1000$. The block size $d$ is set to $5$ in all cases, except in the middle plot where $d=1,2,5,10$, which corresponds to the true value $k_2(\mathbf{x})=10,5,2,1$ (here $k_2(\mathbf{x})$ also equals $\lVert\mathbf{x}\rVert_{2,0}$, the exact block sparsity level of our signal $\mathbf{x}$ with the block size set to $d$). In all cases, $\sigma=0.1$, except in the bottom plot where $\sigma=0,0.1,0.2,0.3$. 

The quantities $|\frac{\hat{k}_2(\mathbf{x})}{k_2(\mathbf{x})}-1|$ are averaged over the $200$ replications as an approximation of $E|\frac{\hat{k}_2(\mathbf{x})}{k_2(\mathbf{x})}-1|$. According to Theorem 2, we get $|\frac{\hat{k}_2(\mathbf{x})}{k_2(\mathbf{x})}-1|\approx \frac{\sqrt{\omega_2}}{\sqrt{n_1+n_2}}|Z|$, where $Z$ is a standard normal random variable and $\omega_2=\frac{\theta_2(c_2,\rho_2)}{\pi_2}+4\frac{\theta_1(c_1,\rho_1)}{1-\pi_2}$. Then the fact that $E|Z|=\sqrt{2/\pi}$ implies that the theoretical curves for $E|\frac{\hat{k}_2(\mathbf{x})}{k_2(\mathbf{x})}-1|$ are simply $\frac{\sqrt{2\omega_2/\pi}}{\sqrt{n_1+n_2}}$, as a function of $n_1+n_2$. It should be note that $\omega_2$ depends on $\sigma$ and $d$, but not on $N$. 

From Figure \ref{fig_3}, we can see that the theoretical curves (black) coincide well with the empirical ones (colored). Moreover, the averaged relative error does not depend on $N$ or $d$ (when $\sigma$ is fixed), as expected from Theorem 2, and its dependence on the $\sigma$ is also negligible. 

b) Meanwhile, the second simulation study in this part has been conducted to illustrate the asymptotic normality of our estimators in Corollary 1 and Theorem 2. We have $1000$ replications for these experiments, that is, we have $1000$ samples of the standardized statistics $\mathrm{res1}=\sqrt{\frac{n_1}{\theta_1(\hat{c}_1,\hat{\rho}_1)}}\left(\frac{\hat{v}_{1}(\hat{t}_{\mathrm{pilot}})}{\lVert\mathbf{x}\rVert_{2,1}}-1\right)$, $\mathrm{res2}=\sqrt{\frac{n_2}{\theta_2(\hat{c}_2,\hat{\rho}_2)}}\left(\frac{\hat{v}_{2}(\hat{t}_{\mathrm{pilot}})}{\lVert\mathbf{x}\rVert_{2,2}^{2}}-1\right)$ and $\mathrm{res}=\sqrt{\frac{n_1+n_2}{\hat{w}_2}}\left(\frac{\hat{k}_2(\mathbf{x})}{k_{2}(\mathbf{x})}-1\right)$. We consider four cases, with $(n_1,n_2)=(500,500), (1000,1000)$ and the noise distribution is standard normal and $t(2)$ which has infinite variance. In all these cases, we fix $N=1000$, $d=5$, and $\sigma=0.1$.

\begin{figure}[!t]
	\centering
	\includegraphics[width=\textwidth,height=0.7\textheight,keepaspectratio]{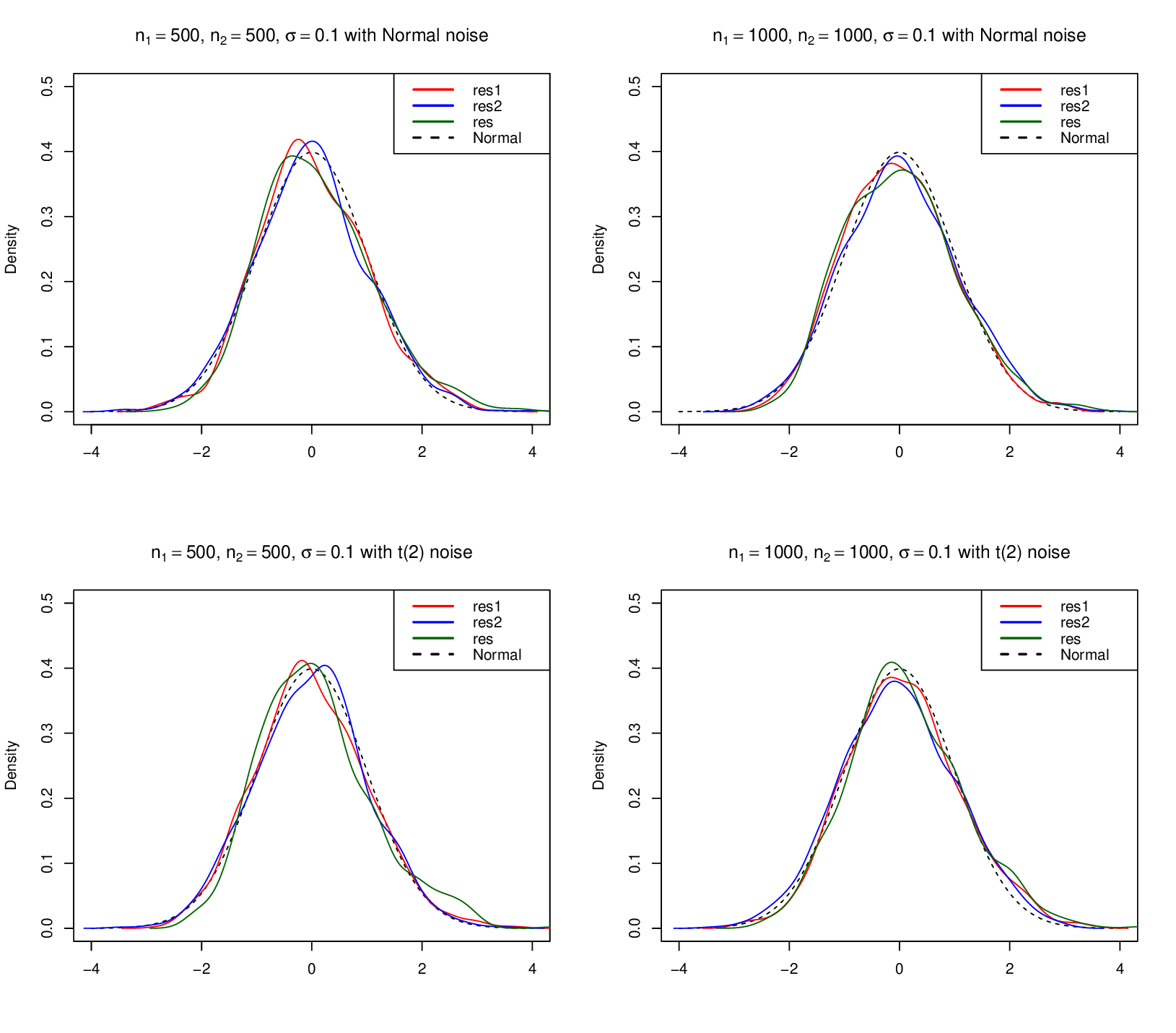}
	\caption{The density plots of the stanardized statistics for the exactly block sparse case. The dashed black curve is the standard normal density in all four plots.}
	\label{fig_4}
\end{figure}

Figure \ref{fig_4} shows that all the probability density curves of the standardized statistics are very close to the standard normal density curve, which verifies our theoretical results. And these results hold even when the noise distribution is heavy-tailed. Comparing these four plots, it can be observed that the normal approximation improves as the sample size $n_1+n_2$ increases or the noise variance decreases.

\subsection{Nearly Block Sparse Case}

\begin{figure}[!t]
	\centering
	\includegraphics[width=\textwidth,height=0.7\textheight,keepaspectratio]{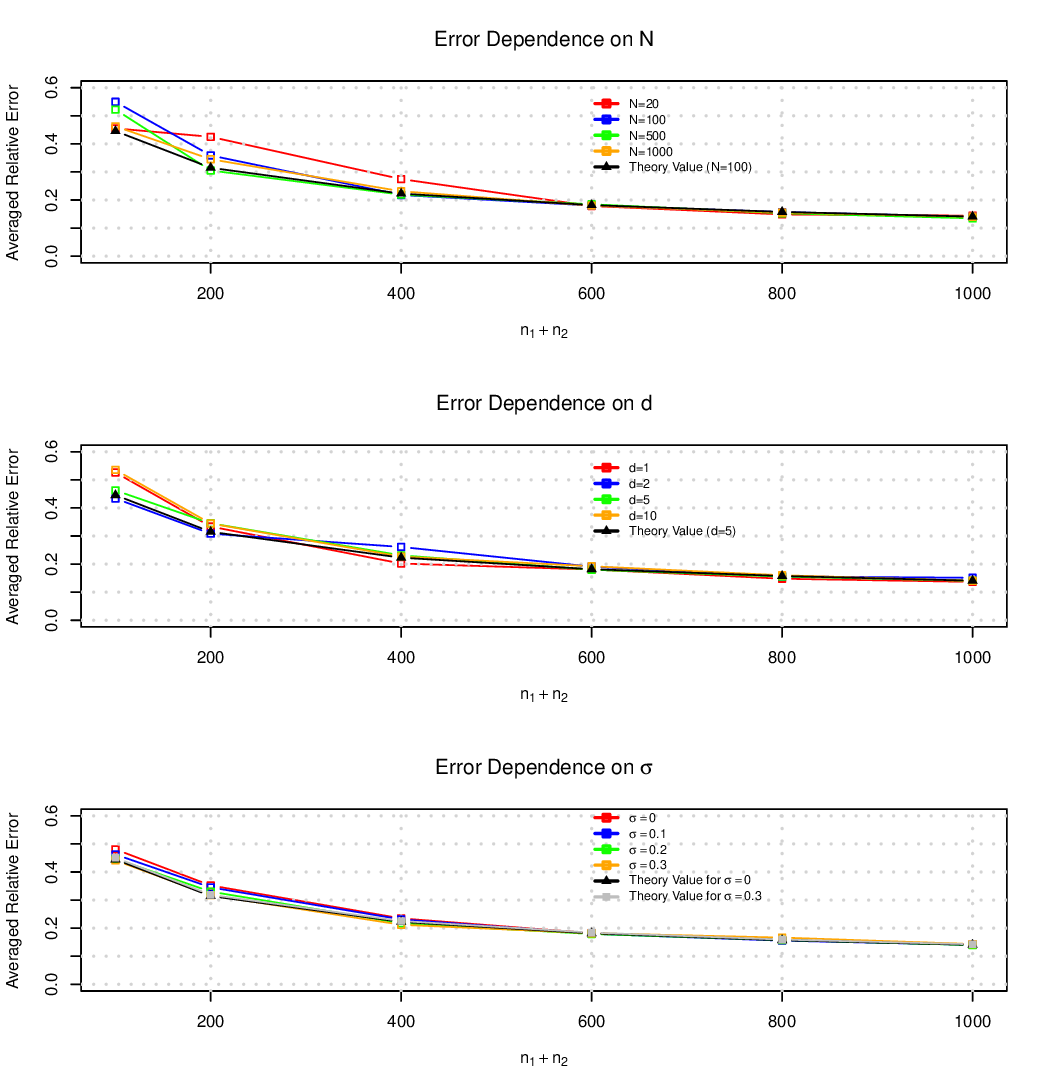}
	\caption{The averaged relative error $|\frac{\hat{k}_2(\mathbf{x})}{k_2(\mathbf{x})}-1|$ depending on $N$, $d$ and $\sigma$ for the nearly block sparse case.}
	\label{fig_5}
\end{figure}

\begin{figure}[!t]
	\centering
	\includegraphics[width=\textwidth,height=0.7\textheight]{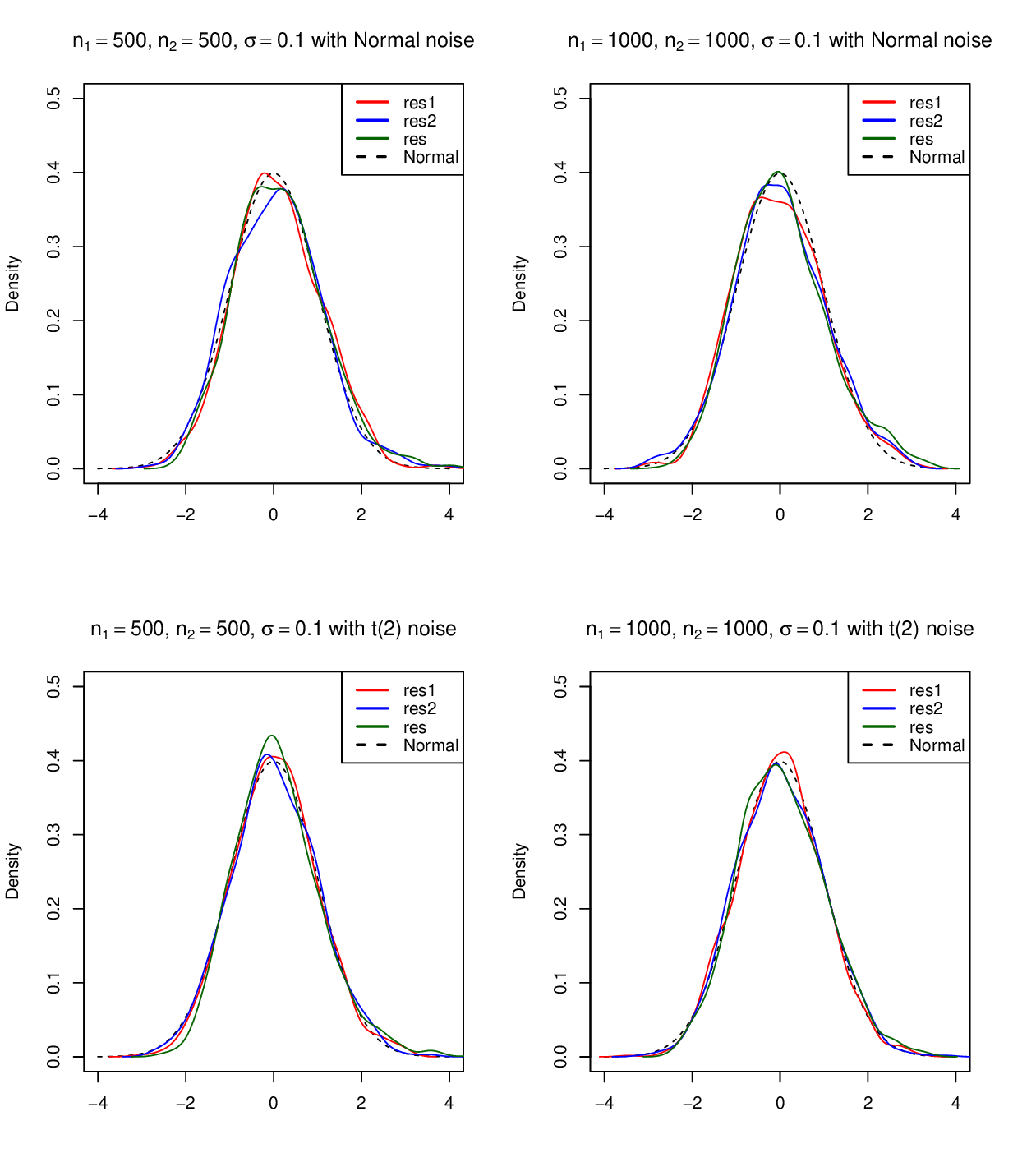}
	\caption{The density plots of the stanardized statistics for the nearly block sparse case. The dashed black curve is the standard normal density in all four plots.}
	\label{fig_6}
\end{figure}

Second, we consider our signal $\mathbf{x}$ to be not exactly block sparse but nearly block sparse, i.e., the entries of $j$-th block all equal $\frac{c}{\sqrt{d}}\cdot j^{-1}$, with $c$ chosen so that $\lVert \mathbf{x}\rVert_{2,2}=\lVert \mathbf{x}\rVert_{2}=1$. In this case, the $\ell_2$ norms of the signal's blocks decay like $j^{-1}$ for $j\in \{1,\cdots,p\}$. Figure \ref{fig_5} and Figure \ref{fig_6} display simulation results that are similar to those obtained in the exactly block sparse case, using the same settings as in the previous subsection.

\subsection{Estimating $\lVert\cdot\rVert_{2,0}$ by $\hat{k}_{\alpha}(\cdot)$ with Small $\alpha$}

Third, we consider the estimation of the mixed $\ell_2/\ell_0$ norm $\lVert\mathbf{x}\rVert_{2,0}$ by using $\hat{k}_{\alpha}(\mathbf{x})$ with $\alpha=0.06$. We consider the signal $\mathbf{x}\in\mathbb{R}^N$ of the form $$c'(\underbrace{\frac{1}{\sqrt{d}}\cdots\frac{1}{\sqrt{d}}}_{d}\underbrace{\frac{1/\sqrt{d}}{2}\cdots\frac{1/\sqrt{d}}{2}}_{d}\cdots
\underbrace{\frac{1/\sqrt{d}}{\lVert\mathbf{x}\rVert_{2,0}}\cdots\frac{1/\sqrt{d}}{\lVert\mathbf{x}\rVert_{2,0}}}_{d}0\cdots0)
$$
with $c'$ chosen so that $\lVert\mathbf{x}\rVert_{2,2}=\lVert\mathbf{x}\rVert_2=1$. In this set of experiments, we fix $N=1000$ and $d=5$. To obtain $\hat{k}_{\alpha}(\mathbf{x})$, we generate the samples $\mathbf{y}_1=A_1\mathbf{x}+\sigma\boldsymbol{\varepsilon_1}$ and $\mathbf{y}_{\alpha}=A_{\alpha}\mathbf{x}+\sigma\boldsymbol{\varepsilon_\alpha}$, where $A_1=(\mathbf{a}_1,\cdots,\mathbf{a}_{n_1})\in\mathbb{R}^{n_1\times N}$ with $\mathbf{a}_i\in\mathbb{R}^N$ being i.i.d. random vector, and $\mathbf{a}_i=(\mathbf{a}_{i1}^T,\cdots,\mathbf{a}_{ip}^T)^T$ with $\mathbf{a}_{ij}, j\in\{1,\cdots,p\}$ i.i.d. drawn from $S(d,1,\gamma_1)$ with $\gamma_1=1$. Similarly,  $A_\alpha=(\mathbf{h}_1,\cdots,\mathbf{h}_{n_{\alpha}})\in\mathbb{R}^{n_{\alpha}\times N}$ with  $\mathbf{h}_i\in\mathbb{R}^N$ being i.i.d. random vector, and $\mathbf{h}_i=(\mathbf{h}_{i1}^T,\cdots,\mathbf{h}_{ip}^T)^T$ with $\mathbf{h}_{ij}, j\in\{1,\cdots,p\}$ i.i.d. drawn from $S(d,\alpha,\gamma_\alpha)$ with $\gamma_\alpha=1$. The noise terms $\boldsymbol{\varepsilon_1}$ and $\boldsymbol{\varepsilon_2}$ are generated with i.i.d. entries from a standard normal distribution. The noise level $\sigma$ is set to $0.1$. We consider a sequence of pairs for the sample sizes $(n_1,n_{\alpha})=(50,50),(100,100),(200,200),\cdots,(500,500)$. For each experiment, we replicate $200$ times. Then, we have $200$ realizations of $\hat{k}_\alpha(\mathbf{x})$ for each $(n_1,n_{\alpha})$. We vary $\lVert\mathbf{x}\rVert_{2,0}$ and $\mathrm{BDNR}(\mathbf{x})$, and average the quantities $\left|\frac{\hat{k}_\alpha(\mathbf{x})}{\lVert\mathbf{x}\rVert_{2,0}}-1\right|$. Specifically, we consider the four cases with $\lVert\mathbf{x}\rVert_{2,0}=\mathrm{BDNR}(\mathbf{x})=10,50,100,200$.

\begin{figure}[!t]
	\centering
	\includegraphics[width=\textwidth,height=0.7\textheight]{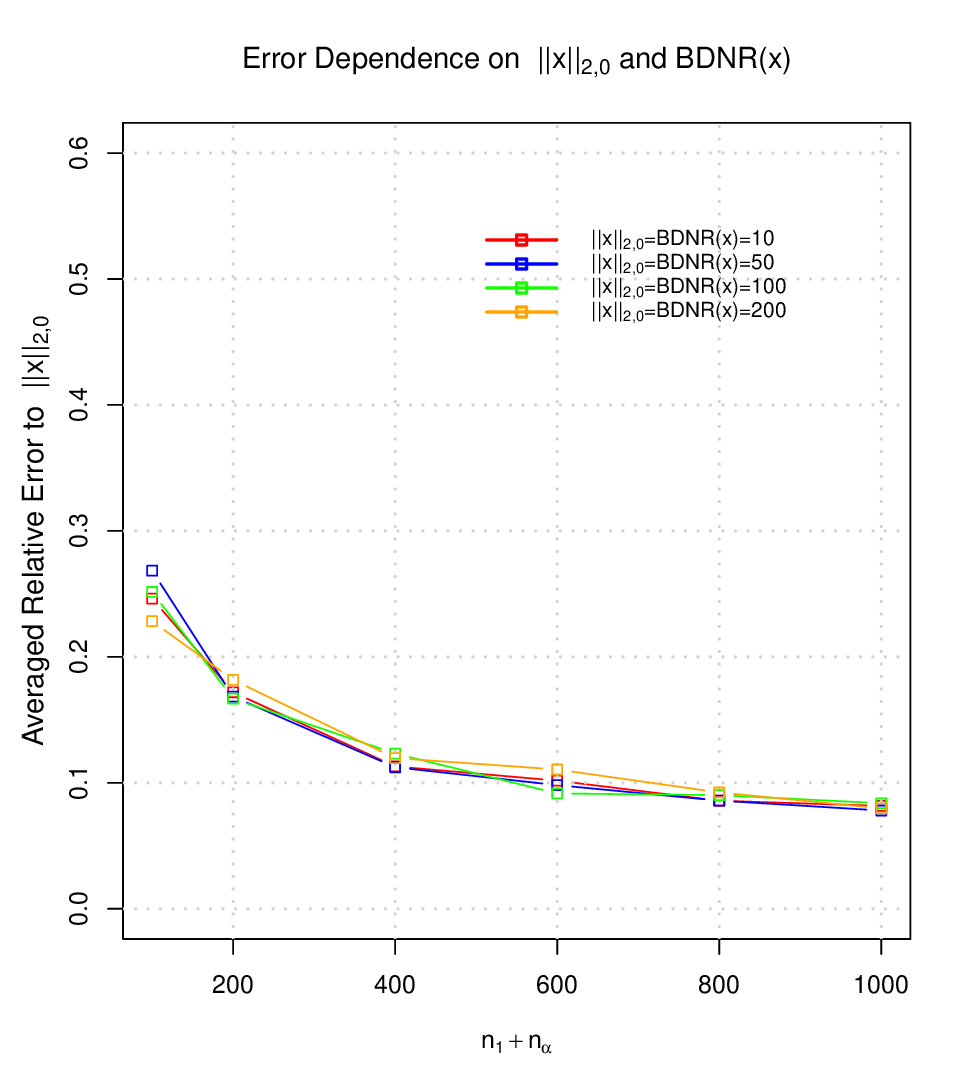}
	\caption{The average relative error $\left|\frac{\hat{k}_{\alpha}(\mathbf{x})}{\lVert\mathbf{x}\rVert_{2,0}}-1\right|$ with $\alpha=0.06$ depending on $\lVert\mathbf{x}\rVert_{2,0}$ and $\mathrm{BDNR}(\mathbf{x})$.}
	\label{fig_7}
\end{figure}

As expected in Theorem 3, it can be seen from Figure \ref{fig_7} that $\hat{k}_{0.06}(\mathbf{x})$ accurately estimates $\lVert\mathbf{x}\rVert_{2,0}$ across a broad set of parameters $\lVert\mathbf{x}\rVert_{2,0}$ and $\mathrm{BDNR}(\mathbf{x})$, and these parameters have a very limited effect on the relative estimation error.

\section{Conclusion}

In this paper, we consider the estimation of block sparsity in compressive sensing, which is crucial from both theoretical and practical points of view. We introduced a new soft measure of block sparsity and developed its estimators by using multivariate centered isotropic symmetric $\alpha$-stable random projections. The asymptotic properties of the estimators were established. A series of numerical experiments illustrated our theoretical results.

There are some interesting issues left for future research. Throughout the paper, we assume that the noise scale parameter $\sigma$ and the characteristic function of noise $\varphi_0$ are known. In practice, however, they are usually unknown and need to be estimated. Although \cite{l2} considered the effects of adopting their estimators in the estimation procedure, how to estimate these parameters based on our random linear projection measurements $\mathbf{y}$ itself is still unknown. In addition, we have been considering the sparsity and block sparsity estimations for real-valued signals so far. It will be interesting to generalize the existing results to the case of complex-valued signals.

\setcounter{secnumdepth}{0}
\section{Appendix\quad Proofs}

Our main theoretical findings Theorem 1 and Theorem 2 follow from Theorem 2 and Corollary 1 in \cite{l2}, because in these two estimation procedures, the noiseless measurements after random projections both have univariate symmetric stable distributions, but with different scale parameters, i.e., $\gamma_\alpha\lVert \mathbf{x}\rVert_{\alpha}$ for the non-block sparsity estimation in \cite{l2} and $\gamma_\alpha\lVert \mathbf{x}\rVert_{2,\alpha}$ for the block sparsity estimation in the present paper. Therefore, the asymptotic results for the scale parameters estimation by using the characteristic function method are rather similar. In order not to repeat, all the details are omitted. Here we only present the proofs of Lemma 3, Lemma 4 and Theorem 3. \\

\noindent
{\bf Proof of Lemma 3.}  The proof procedure follows from the proof of Proposition 1 in \cite{l2} with some careful modifications. We will use the same $c_0$ as in Lemma 2 and some number $\tau_0\geq 1$ that satisfies \begin{align}
\frac{2\ln(\tau_0)+2}{d\tau_0}+\frac{2}{\tau_0}\leq \frac{1}{c_0}.
\end{align}
In Lemma 2, we choose $k=\lfloor t\rfloor$ with $t=\frac{m/\tau_0}{d\ln(\frac{eN}{m})}$. It is worth noting that when $m\leq N$, this $k$ is at most $p$, and thus lies in $\{1,\cdots,p\}$. Then we have \begin{align*}
k\ln(\frac{eN}{kd})&\leq (t+1)\ln(\frac{eN}{td}) \\
&=\left(\frac{m/\tau_0}{d\ln(\frac{eN}{m})}+1\right)\cdot \ln\left(\frac{\tau_0 eN}{m}\cdot\ln(\frac{eN}{m})\right)\\
&\leq \left(\frac{m/\tau_0}{d\ln(\frac{eN}{m})}+1\right)\cdot\ln\left[(\tau_0\frac{eN}{m})^2\right]\\
&=\frac{2m/\tau_0}{d\ln(\frac{eN}{m})}\left(\ln(\tau_0)+\ln(\frac{eN}{m})\right)+2\ln(\tau_0\frac{eN}{m})\\
&\leq\left(\frac{2\ln(\tau_0)+2}{d\tau_0}+\frac{2}{\tau_0}\right)m\leq \frac{m}{c_0},
\end{align*}
by using our assumption that $N\geq m\geq \tau_0\ln(\tau_0\frac{eN}{m})$. Therefore, it is easy to verify that our choice of $\tau_0$ fulfills $m\geq c_0k\ln(eN/kd)$. 

Next, we let $\tau_1=c_1$ be as in Lemma 2 so that (\ref{2.2}) holds with probability at least $1-2\exp(-\tau_1 m)$. Moreover, it holds that $$
\frac{1}{\sqrt{k}}\lVert\mathbf{x}-\mathbf{x}^k\rVert_{2,1}\leq \frac{1}{t}\lVert\mathbf{x}\rVert_{2,1}=\frac{\sqrt{\tau_0}}{\sqrt{m}}\sqrt{d\lVert\mathbf{x}\rVert_{2,1}^2\ln(\frac{eN}{m})}.
$$
Thus, with $c_2$ and $c_3$ being the same as in (\ref{2.2}), we get \begin{align*}
\frac{\lVert\hat{\mathbf{x}}-\mathbf{x}\rVert_2}{\lVert\mathbf{x}\rVert_2}&\leq c_2\frac{\lVert\mathbf{x}-\mathbf{x}^k\rVert_{2,1}}{\sqrt{k}\lVert\mathbf{x}\rVert_2}+c_3\frac{\delta}{\lVert\mathbf{x}\rVert_2}\\
&\leq c_2\frac{\sqrt{\tau_0}}{\sqrt{m}}\sqrt{d\frac{\lVert\mathbf{x}\rVert_{2,1}^2}{\lVert\mathbf{x}\rVert_2^2}\ln(\frac{eN}{m})}
+c_3\frac{\delta}{\lVert\mathbf{x}\rVert_2}.
\end{align*}
Finally, the proof is completed by setting $\tau_2=c_2\sqrt{\tau_0}$, $\tau_3=c_3$ and noticing the fact that $\lVert\mathbf{x}\rVert_2=\lVert\mathbf{x}\rVert_{2,2}$.\\ 

\noindent
{\bf Proof of Lemma 4.} By using the independence of $\mathbf{a}_{1j}, j\in\{1,\cdots,p\}$, for $t\in \mathbb{R}$, the characteristic function of $\langle \mathbf{a}_1,\mathbf{x}\rangle$ has the form: \begin{align*}
E[\exp(\mathrm{i}t\langle \mathbf{a}_1,\mathbf{x}\rangle)]&=E\left[\exp\Big(\mathrm{i}t(\sum_{j=1}^p\mathbf{x}[j]^T\mathbf{a}_{1j})\Big)\right]\\
&=\prod_{j=1}^p E[\exp(\mathrm{i}t\mathbf{x}[j]^T\mathbf{a}_{1j})]\\
&=\prod_{j=1}^p \exp(-\gamma_\alpha^{\alpha}\lVert t\mathbf{x}[j]\rVert_2^\alpha)\\
&=\exp\left[-\gamma_\alpha^{\alpha}\left(\sum_{j=1}^p\lVert\mathbf{x}[j]\rVert_2^\alpha\right)|t|^{\alpha}\right]\\
&=\exp(-(\gamma_\alpha\lVert\mathbf{x}\rVert_{2,\alpha})^\alpha|t|^\alpha).
\end{align*}
Then, this lemma follows from Definition 2.\\

\noindent
{\bf Proof of Theorem 3.} The proof technique of this theorem is adopted from the proof of Proposition 5 in \cite{l2}. We reproduce the proof procedure for the sake of completeness. Combining the triangle inequality that  $$
\frac{|\tilde{k}_{\alpha}(\mathbf{x})-\lVert\mathbf{x}\rVert_{2,0}|}{\lVert\mathbf{x}\rVert_{2,0}}\leq \frac{|\tilde{k}_{\alpha}(\mathbf{x})-k_{\alpha}(\mathbf{x})|}{\lVert\mathbf{x}\rVert_{2,0}}+\frac{|k_{\alpha}(\mathbf{x})-\lVert\mathbf{x}\rVert_{2,0}|}{\lVert\mathbf{x}\rVert_{2,0}}.
$$
and the fact that $k_{\alpha}(\mathbf{x})\leq \lVert\mathbf{x}\rVert_{2,0}$, we get  $$
\left|\frac{\tilde{k}_{\alpha}(\mathbf{x})}{\lVert\mathbf{x}\rVert_{2,0}}-1\right|\leq \left|\frac{\tilde{k}_{\alpha}(\mathbf{x})}{k_{\alpha}(\mathbf{x})}-1\right|+\frac{|k_{\alpha}(\mathbf{x})-\lVert\mathbf{x}\rVert_{2,0}|}{\lVert\mathbf{x}\rVert_{2,0}}.
$$

Thus, to prove Theorem 3, it suffices to bound the last term on the right-hand side. Next, according to the fact that $k_0(\mathbf{x})= \lVert\mathbf{x}\rVert_{2,0}$ and that $k_{\alpha}(\mathbf{x})$ is a non-increasing function of $\alpha$, it follows that $$
|k_{\alpha}(\mathbf{x})-\lVert\mathbf{x}\rVert_{2,0}|=\int_{0}^{\alpha}\left|\frac{d}{du}k_{u}(\mathbf{x})\right|du.
$$

Our task now is to derive a bound on $\left|\frac{d}{du}k_{u}(\mathbf{x})\right|$. For any $j\in\{1,\cdots,p\}$, $u\in(0,\alpha]$ and $\alpha\in (0,1)$, let us introduce the probability vector $\omega_j (\mathbf{x})=\frac{\pi_j(\mathbf{x})^u}{\lVert\pi(\mathbf{x})\rVert_{u}^u}$ with $\pi_j(\mathbf{x})$ and $\pi(\mathbf{x})$ defined as in Section 2.1. Then, according to the definition of $k_u(\cdot)$, we infer that  \begin{align*}
\left|\frac{d}{du}k_{u}(\mathbf{x})\right|&=-\frac{d}{du}k_{u}(\mathbf{x})\\
&=-\frac{d}{du}\exp(H_{u}(\pi(\mathbf{x})))\\
&=-k_{u}(\mathbf{x})\frac{d}{du}H_{u}(\pi(\mathbf{x}))\\
&=-k_{u}(\mathbf{x})\bigg(\frac{-1}{(1-u)^2}\sum\limits_{j:\lVert\mathbf{x}[j]\rVert_2\neq 0}\omega_j(\mathbf{x})\ln\Big(\frac{\omega_j(\mathbf{x})}{\pi_{j}(\mathbf{x})}\Big)\bigg)\\
&\leq \frac{\lVert\mathbf{x}\rVert_{2,0}}{(1-u)^2}\sum\limits_{j:\lVert\mathbf{x}[j]\rVert_2\neq 0}\omega_j(\mathbf{x})\ln\Big(\frac{\omega_j(\mathbf{x})}{\pi_{j}(\mathbf{x})}\Big),
\end{align*}
provided that $k_{u}(\mathbf{x})\leq\lVert\mathbf{x}\rVert_{2,0}$ and the formula for $\frac{d}{du}H_{u}(\pi(\mathbf{x}))$ given in the book \cite{beck1995thermodynamics} (page 72). Furthermore, since $\frac{\omega_j(\mathbf{x})}{\pi_{j}(\mathbf{x})}=\frac{\pi_j(\mathbf{x})^{u-1}}{k_{u}(\mathbf{x})^{1-u}}$ and $k_{\infty}(\mathbf{x})=\frac{\lVert\mathbf{x}\rVert_{2,1}}{\lVert\mathbf{x}\rVert_{2,\infty}}\leq k_{u}(\mathbf{x})\leq \lVert\mathbf{x}\rVert_{2,0}$, then for any $j$ with $\lVert\mathbf{x}[j]\rVert_2\neq 0$, we have \begin{align*}
\frac{\omega_{j}(\mathbf{x})}{\pi_{j}(\mathbf{x})}\leq \frac{\pi_{j}(\mathbf{x})^{u-1}}{k_{\infty}(\mathbf{x})^{1-u}}
&\leq \frac{\pi_{j}(\mathbf{x})^{-1}}{k_{\infty}(\mathbf{x})}\cdot k_{\infty}(\mathbf{x})^u \\
&=\frac{\lVert\mathbf{x}\rVert_{2,1}}{\lVert\mathbf{x}[j]\rVert_2}\frac{\lVert\mathbf{x}\rVert_{2,\infty}}{\lVert\mathbf{x}\rVert_{2,1}}\cdot k_{\infty}(\mathbf{x})^u \\
&\leq \mathrm{BDNR}(\mathbf{x})\cdot\lVert\mathbf{x}\rVert_{2,0}^u,
\end{align*}
where $\mathrm{BDNR}(\mathbf{x})=\frac{\lVert\mathbf{x}\rVert_{2,\infty}}{|\mathbf{x}|_{2,\mathrm{min}}}$. As a result, 
for all $u\in (0,\alpha]$ and $\alpha\in (0,1)$, it holds that \begin{align*}
\left|\frac{d}{du}k_{u}(\mathbf{x})\right|&\leq \frac{\lVert\mathbf{x}\rVert_{2,0}}{(1-u)^2}\ln\left( \mathrm{BDNR}(\mathbf{x})\cdot\lVert\mathbf{x}\rVert_{2,0}^u\right)\sum\limits_{j:\lVert\mathbf{x}[j]\rVert_2\neq 0}\omega_j(\mathbf{x}) \\
&\leq \frac{\lVert\mathbf{x}\rVert_{2,0}}{(1-u)^2}\Big(\ln(\mathrm{BDNR}(\mathbf{x}))+u\ln(\lVert\mathbf{x}\rVert_{2,0})\Big) \\
&\leq\frac{\lVert\mathbf{x}\rVert_{2,0}}{(1-u)^2}\Big(\ln(\mathrm{BDNR}(\mathbf{x}))+\alpha\ln(\lVert\mathbf{x}\rVert_{2,0})\Big),
\end{align*}
due to $\sum\limits_{j:\lVert\mathbf{x}[j]\rVert_2\neq 0}\omega_j(\mathbf{x})\leq \sum\limits_{j=1}^p \omega_j(\mathbf{x})=1$. Hence, the proof is completed according to the basic integral result $\int_0^{\alpha}\frac{1}{(1-u)^2}du=\frac{\alpha}{1-\alpha}$.

\bibliographystyle{spmpsci}
\bibliography{blocksparsity}


%
%


%
%

\end{document}